\documentclass[iop]{emulateapj}
\newcommand{\mytilde}{\raise.19ex\hbox{$\scriptstyle\sim$}}
\usepackage{amsmath}
\usepackage{xcolor}
\usepackage{soul}
\usepackage{upgreek}
\graphicspath{{./}{figures/}}
\usepackage[utf8]{inputenc}
\shorttitle{WL Mass Bias in Mergers}
\shortauthors{Lee et al.}
\begin{document}

\title{Weak-lensing mass bias in merging galaxy clusters}

\author{Wonki Lee\altaffilmark{1,2}}
\author{Sangjun Cha\altaffilmark{1}}
\author{M. James Jee\altaffilmark{1,3}}
\author{Daisuke Nagai\altaffilmark{4,5}}
\author{Lindsay King\altaffilmark{6}}
\author{John ZuHone\altaffilmark{2}}
\author{Urmila Chadayammuri\altaffilmark{2}}
\author{Sharon Felix\altaffilmark{6}}
\author{Kyle Finner\altaffilmark{7}}

\altaffiltext{1}{Yonsei University, Department of Astronomy, Seoul, Republic of Korea; wonki.lee@yonsei.ac.kr, mkjee@yonsei.ac.kr}
\altaffiltext{2}{Harvard-Smithsonian Center for Astrophysics, 60 Garden St., Cambridge, MA 02138, USA}
\altaffiltext{3}{Department of Physics, University of California, Davis, One Shields Avenue, Davis, CA 95616, USA}
\altaffiltext{4}{Department of Physics, Yale University, New Haven, CT 06520, USA}
\altaffiltext{5}{Department of Astronomy, Yale University, New Haven, CT 06511, USA}
\altaffiltext{6}{Department of Physics, University of Texas - Dallas, Richardson, TX 75080, USA}
\altaffiltext{7}{Infrared Processing and Analysis Center, California Institute of Technology, 1200 E California Blvd., Pasadena, CA 91125, USA}

\begin{abstract}
Although weak lensing (WL) is a powerful method to estimate a galaxy cluster mass without any dynamical assumptions, a model bias can arise when the cluster density profile departs from the assumed model profile. In a merging system, the bias is expected to become most severe because the constituent halos undergo significant structural changes. In this study, we investigate WL mass bias in binary cluster mergers using a suite of idealized hydrodynamical simulations. Realistic WL shear catalogs are generated by matching the source galaxy properties, such as intrinsic shape dispersion, measurement noise, source densities, etc., to those from Subaru and {\it Hubble Space Telescope} observations. We find that, with the typical mass-concentration ($M$-$c$) relation and the Navarro-Frenk-White (NFW) profile, the halo mass bias depends on the time since the first pericenter passage and increases with the mass of the companion cluster. The time evolution of the mass bias is similar to that of the concentration, indicating that, to first order, the mass bias is modulated by the concentration change. For a collision between two $\mytilde10^{15}~M_{\odot}$ clusters, the maximum bias amounts to $\mytilde60\%$. This suggests that previous WL studies may have significantly overestimated the mass of the clusters in some of the most massive mergers.
%We show that the WL mass bias due to the $M$-$c$ relation and the NFW assumption can be alleviated when an accurate concentration is independently provided, and the merger phase is estimated and utilized. 
Finally, we apply our results to three merger cases: Abell~2034, MACS~J1752.0+4440, and ZwCl~1856.8+6616, and report their mass biases at the observed epoch, as well as their pre-merger masses, utilizing their merger shock locations as tracers of the merger phases.
\end{abstract}

\section{Introduction} \label{sec:intro}

In the hierarchical structure formation scenario, a merger is one of the major channels through which galaxy clusters grow their mass.
During the collision between clusters, an order of $\mytilde 10^{64} \, \rm erg$ \citep[e.g.,][]{Ricker2001} gravitational energy is transferred to the cluster environment, which creates a unique experimental setup for testing various astrophysical processes. 
This includes energy transport via thermal conduction \citep[e.g.,][]{1999ApJ...521..526M,2003ApJ...586L..19M},  development of the intracluster magnetic field
\citep[e.g.,][]{2005ApJ...627..733M,2011ApJ...743...16Z}, the electron-ion equilibrium timescale \citep[e.g.,][]{2006ESASP.604..723M,2012MNRAS.423..236R}, acceleration of high-beta plasma \citep[e.g.,][]{2007MNRAS.378..245B,2012ApJ...756...97K}, and self-interaction cross-section of dark matter \citep[e.g.,][]{2004ApJ...606..819M,2008ApJ...679.1173R}. 

However, astrophysical constraints based on merging galaxy clusters generally suffer from uncertainties in the merger configuration (such as viewing angle, collision speed, time since the collision, pericenter distance, mass, mass ratio, etc). Thus, it is crucial to robustly characterize the merging scenario to utilize the cluster merger as an astrophysical laboratory.

Mass is one of the most critical parameters that affect the merging scenario. However, determining the mass of a dynamically active
cluster is very challenging.
Mass estimators that use the properties of thermalized plasma from X-ray \citep[e.g.,][]{2006ApJ...640..691V} or Sunyaev-Zel’dovich effect \citep[SZ,][for a recent review]{1970Ap&SS...7....3S,1972CoASP...4..173S,Carlstrom2002,Mroczkowski2019}  
observations are biased since the condition of hydrostatic equilibrium (HSE) does not hold during the merger \citep[e.g.,][]{Nagai2007b,2012MNRAS.419.1766K, Nelson2012,Nelson2014,Shi2016}.
In this sense, weak-lensing (WL) is a promising tool because its signal is only sensitive to the projected potential, not depending on the dynamical properties of the system. 
This unique capability has been used to map complex substructures in merging clusters
\citep[e.g.,][]{Hoekstra2000, Clowe2006, okabe2008, 2016ApJ...817..179J, Kim2019, yoon2020}, which
aid numerical simulations by providing constraints on the amount and location of the subcluster halos \citep[e.g.,][]{Springel2007,Milosavljevic2007,Lage2014,Molnar2015, 2018ApJ...862..160W,2018ApJ...855...36Z,2020ApJ...894...60L,2022MNRAS.509.1201C}.

The weakness of WL, however, is the so-called mass-sheet degeneracy, where the signal is invariant under the transformation of the surface mass density: $\kappa \rightarrow \lambda \kappa + 1 - \lambda$ ($\kappa$: dimensionless mass density, $\lambda$: free parameter, described in more details in \textsection\ref{sec:mockWL_method}). Therefore, WL practitioners generally estimate cluster masses by assuming that the halos follow analytic profiles. The most popular choice among many analytic models is the Navarro-Frenk-White \citep[NFW,][]{Navarro1996} model with
the assumption that the mass correlates with the halo concentration \citep[e.g.,][]{Duffy2008}.
Since both the NFW profile and the mass-concentration ($M$-$c$) relation are valid only for the description of the average halo properties \citep[e.g.,][]{2000ApJ...535...30J}, systematic errors are inevitable when the method is applied to individual clusters.
Previous studies have discussed both scatter and bias from these assumptions, showing that the triaxial shape of the dark halo \citep[e.g.,][]{2004MNRAS.350.1038C}, projected structures along the line-of-sight \citep[e.g.,][]{2011MNRAS.412.2095H}, miscentering \citep[e.g.,][]{2007arXiv0709.1159J}, and halo concentration \citep[e.g.,][]{2018MNRAS.474.2635S} can cause
uncertainties at the $10-50\%$ level.

The departure from the NFW profile and $M$-$c$ relation is most pronounced in merging clusters, where two systems severely disrupt each other.
Numerical studies have suggested that the halo compactness can sharply increase after the first closest passage \citep[][]{2012MNRAS.419.1338R,2021arXiv211204926R,2022MNRAS.509.1201C}.
Observational studies showed that merging clusters can have either higher \citep[e.g.,][]{2008ApJ...685L...9B} or lower \citep[e.g.,][]{2019PASJ...71...79O} concentration values than the $M$-$c$ relation prediction. \citet{2022MNRAS.509.1201C} reported that in the case of Abell 2146, the WL mass could be greater than the pre-merger mass by $\mytilde2$ when the merger-driven concentration change is not accounted for in the model fitting.
However, no detailed studies have been conducted to quantify the merging cluster WL mass bias for a wide range of masses and mass ratios at various merger phases.

In this study, we use a suite of idealized simulations of cluster mergers to understand the bias of WL mass estimation in post-mergers when the cluster masses are derived by the conventional method based on the NFW fitting with the $M$-$c$ relation.
To mimic an observational procedure, we generate mock WL catalogs by matching the source galaxy properties, such as intrinsic shape dispersion, measurement noise, source densities, etc., to those from real telescope data. 

This paper is organized as follows. 
The setups for the idealized cluster merger simulations and the mock WL analysis are described in \textsection\ref{sec:Sim}. 
We present the time evolution of concentration in \textsection\ref{sec:res}, the bias of WL mass estimates in \textsection\ref{sec:WLmass}, and the application to previous observation studies in \textsection\ref{sec:obs}. 
Discussion is presented in \textsection\ref{sec:discussion} before the conclusions in \textsection\ref{sec:conclusions}. 
We adopt a $\Lambda$CDM cosmology with $H_0=70$ km~s$^{-1}$Mpc$^{-1}$, $\Omega_m=0.3$, and $\Omega_{\Lambda}=0.7$. Throughout the paper, we assume that the clusters are at $z=0.2$. 
$R_{\Delta}$ describes the radius where the average density becomes $\Delta$ times the critical density of the universe. 
$M_{\Delta}$ is the total mass within $R_{\Delta}$.

\section{Simulations} \label{sec:Sim}
We use \texttt{GAMER-2} \citep[][]{2018MNRAS.481.4815S}, a GPU-accelerated Adaptive MEsh Refinement code, to perform idealized hydrodynamical simulations of merging clusters.
We solve Euler equations using the MUSCL-Hancock method \citep[MHM,][]{doi:10.1137/0905001}, the HLLC scheme \citep[][]{Toro1994}, and implement a van Leer-type limiter in the slope computation.
We do not implement radiative physics, including cooling and baryonic feedback, as we mainly focus on the dark halo properties %in
on the scale greater than $0.15\,r_{500}$ \citep[e.g.,][]{2006ApJ...650..128K}.
The simulation refines each sub-region of interest by defining a ``patch'' (also known as a ``grid'' or ``block''). Each patch consists of a fixed number of cells ($8^3$), and the refinement is triggered when
the number of dark matter particles exceeds a threshold.
This allows us to adaptively refine our merging cluster inside a $15\rm~Mpc$-side box with the sparsest resolution of $\mytilde200\rm~kpc$ in the outskirts to the finest level $\mytilde15\rm~kpc$ at the cluster center. 
%We confirm that our results are consistent with the $\mytilde7\rm~kpc$-resolution simulation results.
We performed a resolution test by running simulations with the finest cell size of $\mytilde7\rm~kpc$ and found that the results, including the halo concentration and the WL mass estimate, are consistent with the $\mytilde15\rm~kpc$-resolution results.

\subsection{Simulation setups} \label{sec:Ref_IC}

For the description of the impact of various collision parameters on the halo concentration and the WL mass estimation, we define a reference run.
The reference run is a collision between $5\times10^{14}\,M_{\odot}$ and $1\times10^{14}\,M_{\odot}$ clusters.  We set the baryon fraction to $0.13$ at $R_{200}$ based on the cosmic baryon fraction of \citet{2013ApJS..208...19H} and the stellar mass fraction of \citet{2009ApJ...703..982G}.
The initial distribution of dark matter particles and ICM profiles is generated using the \texttt{cluster\_generator} \footnote{https://github.com/jzuhone/cluster\_generator} package.
The position of dark matter particles is randomly generated based on the input 
mass profile. Then, we calculate the velocity of each particle at its
position using the Eddington Formula \citep{1916MNRAS..76..572E}. 
Similarly, for the ICM, we derive a pressure profile that satisfies hydrostatic equilibrium. 
The gas properties of each cell are interpolated based on the derived pressure profile and the assigned gas density profile.  
%The cluster profiles are defined up to a radius of $5\rm~Mpc$. For the cells located within $5\rm~Mpc$ from both cluster centers, we add gas densities, momentums, and pressures to populate the cell properties. The background gas density and temperature are set to $\mytilde 10^{-30}\,$g\,cm$^{-3}$ and $\mytilde 10^{7}\,$K, respectively.

We let the dark halo follow an NFW profile, which we also assumed in our mock WL analysis.
The density is defined as
\begin{equation}
\label{eq:dmhalo}
\rho_{\rm DM} (r) =\frac{200c^3 g(c)\rho_c}{3(r/r_{\rm s})\,(1+r/r_{\rm s})^{2}},
\end{equation}
where $r_{\rm s}$ is the scale radius, $c$ is the concentration ($c=r_{200}/r_{\rm s}$), and  $g(c)=[\ln(1+c)-c/(1+c)]^{-1}$ \citep{Navarro1996}.
We use the $M$-$c$ relation from \citet{Duffy2008} and
adopt the coefficients derived at $z=0$.
We vary the initial concentrations within $c=[2.5,5.5]$
so that our study explores potential systematics due to the different coefficients in the $M$-$c$ relations \citep[e.g.,][]{2015ApJ...799..108D}.

We model the ICM density with the modified beta profile suggested by \citet{2006ApJ...640..710V}. These profiles are defined as
\begin{equation}
n_{\rm p}n_{\rm e}=n^2_0 \frac{(r/r_{c})^{-\alpha}}{(1+r^2/r^2_{\rm c})^{3\beta-\alpha/2}}\frac{1}{(1+r^\gamma/r^\gamma_{\rm s,g})^{\epsilon/\gamma}},
\end{equation}
where $n_{\rm p}$ and $n_{\rm e}$ are the number densities of the protons and electrons,
respectively, 
$r_{\rm c}$ is the core radius, $r_{\rm s,g}$ is the scale radius of the gaseous halo, and $\alpha,\,\beta,\,\gamma,\epsilon$ describe the slope of the gas density. 
We set $r_{\rm c}$ ($r_{\rm s,g}$) as $0.2\times r_{2500}$ ($0.67\times r_{200}$). 
The slope parameters are set to $(\alpha, \beta, \gamma,\epsilon)=(1.0,\,0.67,\,3.0,\,3.0)$, which generates a cool-core cluster.

We place the two clusters at an initial separation of $3\rm~Mpc$, which is larger than the sum of the two $R_{200}$ values. 
The cluster profiles are defined up to a radius of $5\rm~Mpc$. 
For the cells located within $5\rm~Mpc$ from both cluster centers, we add the gas densities, momentums, and pressures to populate the cell properties. The background gas density and temperature are set to $\mytilde 10^{-30}\,$g\,cm$^{-3}$ and $\mytilde 10^{7}\,$K, respectively.
The initial in-fall velocity is set as $1200\rm~km~s^{-1}$, which is a factor of $\mytilde1.3$ greater than the circular velocity at the initial separation. 
The reference run is set as a head-on cluster merger (i.e., zero impact parameter) in
the plane of the sky. 
Hereafter, we refer to the more (less) massive cluster as the main (sub) cluster.

\subsubsection{Variation of collision parameters and cluster masses}
\label{sec:collision_parameters}
We test the dependence of the three collision parameters: impact parameter, velocity, and  concentration by varying one parameter at a time from the reference run.
The impact parameter ranges from $0$ (i.e., head-on collision) to $1\rm~Mpc$, which results in the pericenter distance (the separation at the first closest passage) ranging from 0 to $\mytilde0.35\rm~Mpc$.
We vary the initial velocity from $600\rm~km\,s^{-1}$ to $1800\rm~km\,s^{-1}$, which
is comparable to the escape velocity. This translates to a range in the collision velocity $3300-3600\rm~km\,s^{-1}$.
The pre-merger concentration of each cluster is varied from $2.5$ to $5.5$. 

Apart from the above, in order to examine the impact of the mass and mass ratios, we vary the mass of the main cluster from $10^{14}$ to $10^{15}M_{\odot}$.
For each mass of the main cluster, the sub-cluster mass is either fixed to $10^{14}M_{\odot}$ or modified in such a way that the mass ratio is maintained to be 5:1.
To create similar collision parameter setups to the reference run,
we modify the impact parameter and initial velocity as follows.
The initial velocity is adjusted to have the same virial ratio as the reference run. 
Also, the impact parameter $b$ is modified so that $b/R_{200}$ remains the same.
As in the reference run, the initial concentration is %set based on the
determined from the $M$-$c$ relation of \citet{Duffy2008}.

\begin{figure}
    \centering
	\includegraphics[width=\columnwidth]{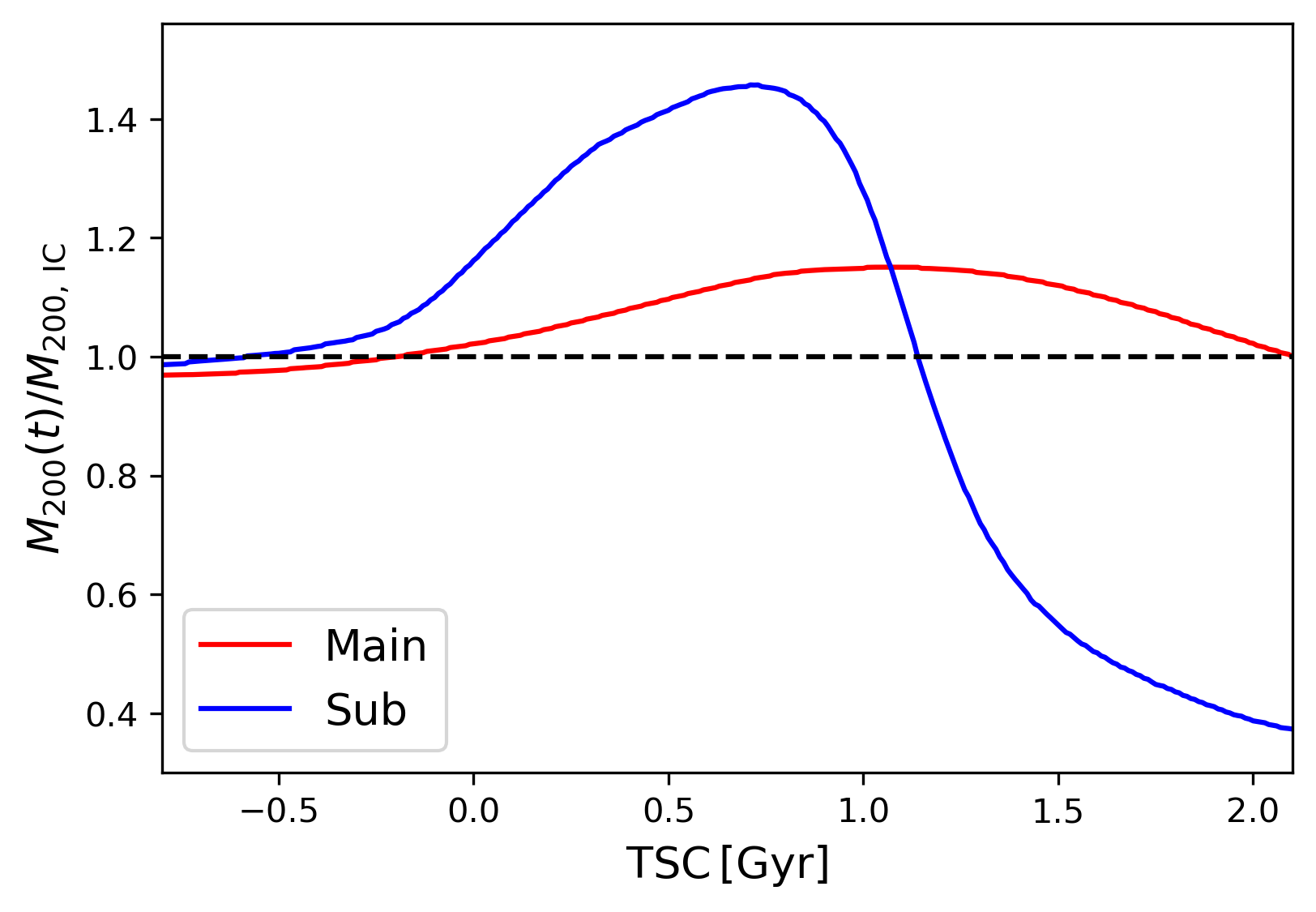}
    \caption{Time evolution of $M_{200}$ in the reference run. 
    We normalize $M_{200}$ at epoch $t$ with its initial (pre-merger) value $M_{200,IC}$.
    The epoch is measured from the time when two halo cores overlap (i.e., TSC).
    After the core passage (TSC$>0$), $M_{200}$ of the clusters evolves in time, which will inflect the WL mass estimate.
    }
    \label{fig:Mvirevol}
\end{figure}

\subsection{Definition of dark halo properties}
Since the main goal of the current study is to track the time evolution of halo
properties and WL mass bias, it is necessary to robustly define the halo center and profile. Here we follow the \citet{2003MNRAS.338...14P} approach.
First, we divide the dark matter particles into two groups based on their membership at the initial condition setup. 
Second, we compute the center of mass of each particle group to make an initial guess of the halo center.
Third, we recursively refine the center for each group using the dark matter particles residing within a shrinking sphere.
At each step, we choose the center of mass defined in the previous step and shrink the radius by $20\%$. 
The initial radius of the shrinking circle is 2~Mpc and the search stops when the value becomes less than $0.2\rm~Mpc$.
We visually verify that the center
determined in this way robustly follows the density peak of the dark matter distribution. 

Once the halo center is determined, we obtain $R_{200}$ and $M_{200}$ from the total (DM+ICM) density profile.
As before, we compute the density using the dark matter particles and the ICM belonging to each group.
For ICM, we define a scalar field, describing each cluster's gas fraction.
Then, we generate ICM density profiles only using the gas mass that belongs to the cluster at the initial condition setup.
However, when it comes to the measurement of the dark halo scale radius $r_s$, we do not fit the model (NFW+modifed-beta profile) on the total density profile as the ICM drags out from the dark halo center during the major merger.
Instead, $r_s$ is calculated by fitting an NFW model (eqn~\ref{eq:dmhalo}) on the dark matter density profile. %\footnote{We find that this model simplification can cause systematic uncertainty of $r_s$ }.
The dark matter density profile is log-binned from $r=10\rm~kpc$ to $3\rm~Mpc$, and Poisson noise is assumed on each bin.
The concentration parameter is estimated from the ratio $R_{200}/r_s$.

%Once the halo center is determined, the dark matter density profile is constructed.
%As before, we compute the density only using the particles belonging to each group.
%The profile is log-binned from $r=10\rm~kpc$ to $3\rm~Mpc$, and Poisson noise is assumed on each bin. 
%The dark halo scale radius $r_s$ is calculated by fitting an NFW model (eqn~\ref{eq:dmhalo}) to this density profile.
%We fit the NFW profile using the DM density as the ICM drags out from the dark halo center during the major merger.
%However, when it comes to the measurement of $R_{200}$, we do not use the NFW fitting result because we need to include both DM and ICM. Thus, we obtain $R_{200}$ and $M_{200}$ from the total (DM+ICM) density profile.
%For ICM, we define a scalar field, describing each cluster's gas fraction.
%Similar to the DM density profile, we generate ICM density profiles only using the gas mass that belongs to the cluster at the initial condition setup.
%The concentration parameter is estimated from the ratio $R_{200}/r_s$.

\subsubsection{Evolution of $M_{200}$ during post-merger}\label{sec:dh_properties}

Figure \ref{fig:Mvirevol} shows the time evolution of $M_{200}$ of the main and sub-clusters
in the reference run. The $M_{200}(t)$ value at each merger stage $t$ is normalized by the initial (pre-merger) mass $M_{200,IC}$. 
The epoch is described with the time since collision (TSC, the time since the two cores overlap). This first encounter (TSC$=0$) happens $\mytilde1.56\rm~Gyrs$ after the initial setup where they are separated by 3~Mpc.
Both cluster masses stay close to their initial values at TSC$<0$ and gradually increase at TSC$>0$.
The main cluster mass reaches its maximum at the first apocenter ($\mytilde1.2\rm~Gyrs$) and slowly decreases afterward. On the other hand, the maximum of the sub-cluster mass occurs sooner ($\mytilde0.8\rm~Gyrs$), and the decrease is faster than its increase.

The difference in the pattern of the mass change between the two clusters is attributed
to the contraction of the halos due to the increased gravitational field from the pericenter passage \citep[e.g.,][]{2012MNRAS.419.1338R}.
%to the different halo contraction triggered by the increased gravity during the halo overlap \citep[e.g.,][]{2012MNRAS.419.1338R}.
One can conjecture that this contraction will also affect the time evolution of the halo concentration in a similar way. A quantitative comparison with the halo contraction will be discussed in \textsection \ref{sec:res}. The sub-cluster mass decrease is much more significant because its shallower potential cannot retain the outflowing mass after the core passage.

\subsection{Mock WL Analysis}\label{sec:mockWL_method}

For our cluster merger simulations (\textsection\ref{sec:Sim}), we carry out a mock WL analysis to investigate how the $M_{200}$ variation translates to WL mass bias. At each merger phase, we generate a synthetic WL ellipticity catalog based on the projected mass distribution and determine the halo masses by fitting two NFW profiles simultaneously.

We represent a $5\rm~Mpc \times 5\rm~Mpc$ projected mass distribution with
a $1000\times1000$ grid.
The projected mass density $\Sigma$ is converted to a convergence $\kappa$, which is the dimensionless surface mass density:
\begin{equation}
\kappa=\frac{\Sigma}{\Sigma_{c}},
\end{equation}
where $\Sigma_{c}$ is the critical surface mass density. 
The critical surface mass density is defined as:
\begin{equation}
\Sigma_{c}=\frac{{c^{2}}{D_{s}}}{4{\pi}G{D_{l}}{D_{ls}}},
\end{equation}
where $c$ is the speed of light, $G$ is the gravitational constant, $D_{l}(D_{s})$ is the angular diameter distance to the cluster (source), and $D_{ls}$ is the angular diameter distance from the cluster to the source. 

We generate mock WL observational data from the converted convergence above \citep{2001PhR...340..291B}. 
First, from the convergence map, we compute a reduced shear $g$ field by the following:
\begin{equation}
g=\frac{\gamma}{1-{\kappa}}.
\end{equation}
The shear $\mathbf{\gamma}$ is calculated by:
\begin{equation}
\boldsymbol{\gamma}(\mathbf{x})=\frac{1}{\pi}\int\mathbf{D}(\mathbf{x}-\mathbf{x}^{\prime}) \kappa(\mathbf{x}^{\prime})d\mathbf{x}^{\prime}, \label{eqn_shear}
\end{equation}
where $\mathbf{D}$ is the convolution kernel:
\begin{equation}
\mathbf{D}=- \frac{1}{ (x_1- i x_2)^2 }.
\end{equation}
Note that the computation of $\gamma$ through equation~\ref{eqn_shear} is in principle biased because the  convergence $\kappa$ outside the field ($5\rm~Mpc \times 5\rm~Mpc$) is ignored (in a typical WL study, 
the extent of the NFW profile is assumed to be infinite).
The bias due to this finite-field effect is noticeable only at $r\gtrsim2.2$~Mpc. In this study, we choose to
exclude the WL data in this regime.
In addition to this finite-field effect, the finite resolution of the grid creates some ($\mytilde10$\%) bias in reduced shear prediction near the cluster peak ($r\lesssim250$~kpc). Hence, we also avoid using the WL data at such small radii. 

The reduced shear $g$ field provides the observed ellipticity variation across the field if the intrinsic shape of the source is a perfect circle.
To mimic the real observational WL data, we need to randomize the intrinsic source ellipticity, measurement noise, and galaxy position.
Under the reduced shear $\mathbf{g}$, the intrinsic source ellipticity $\mathbf{e}$ at the random position $(x,y)$ is modified to the observed ellipticity $\boldsymbol{\epsilon}$ according to the following rules:
\begin{equation}
\boldsymbol{\epsilon}=\frac{\mathbf{e}+\mathbf{g}}{1+{\mathbf{g}}^{*}{\mathbf{e}}} ~      (|\mathbf{g}|<1)
\end{equation}
and
\begin{equation}
\boldsymbol{\epsilon}=\frac{1+\mathbf{g}\mathbf{e}^{*}}{{\mathbf{e}}^{*}+{\mathbf{g}}^{*}} ~    (|\mathbf{g}|>1),
\end{equation} 
where the asterisk ($*$) indicates the complex conjugate. When randomizing $\mathbf{e}$, we assume $\sigma_e=0.25$ per component.
The final source galaxy ellipticity $\boldsymbol{\epsilon}^{'}$ is obtained after we add measurement noise $\sigma_m$ to each component of $\boldsymbol{\epsilon}$. The measurement noise $\sigma_m$ depends on the S/N of a source. We utilize the \emph{HST} WL catalog of \citet{2021ApJ...923..101K} to derive the relation between source S/N and ellipticity measurement error.
We use the following $\chi^2$-minimization to find the best-fit NFW models from the mock WL catalog:
\begin{equation}
\chi^{2}=\sum_{i=1}^{I} \sum_{j=1}^{2} \frac{{(g_j^i}-\epsilon^{'i}_j)^2}{{\sigma_{s,i}}^{2}+{\sigma_{m,i}}^{2}},
\end{equation}
where $I$ is the total number of background source galaxies, 
$\epsilon^{'i}_j$ is the $j^{th}$ component of the noisy ellipticity for the $i^{th}$ source, and
$g_j^i$ is the $j^{th}$ component reduced shear expected
at the location of the $i^{th}$ source.
We set the source redshift to $z_s=1$ and the source density to 100 (25) galaxies$~\rm arcmin^{-2}$ to follow the \emph {HST} (Subaru) WL catalog. At the redshift of the cluster $z=0.2$, the field size corresponds to $25\farcm86\times25\farcm86$.
If not specified, the WL mass estimate assumes the \citet{Duffy2008} $M$-$c$ relation in the NFW model fitting.

\begin{figure}
    \centering
	\includegraphics[width=\columnwidth]{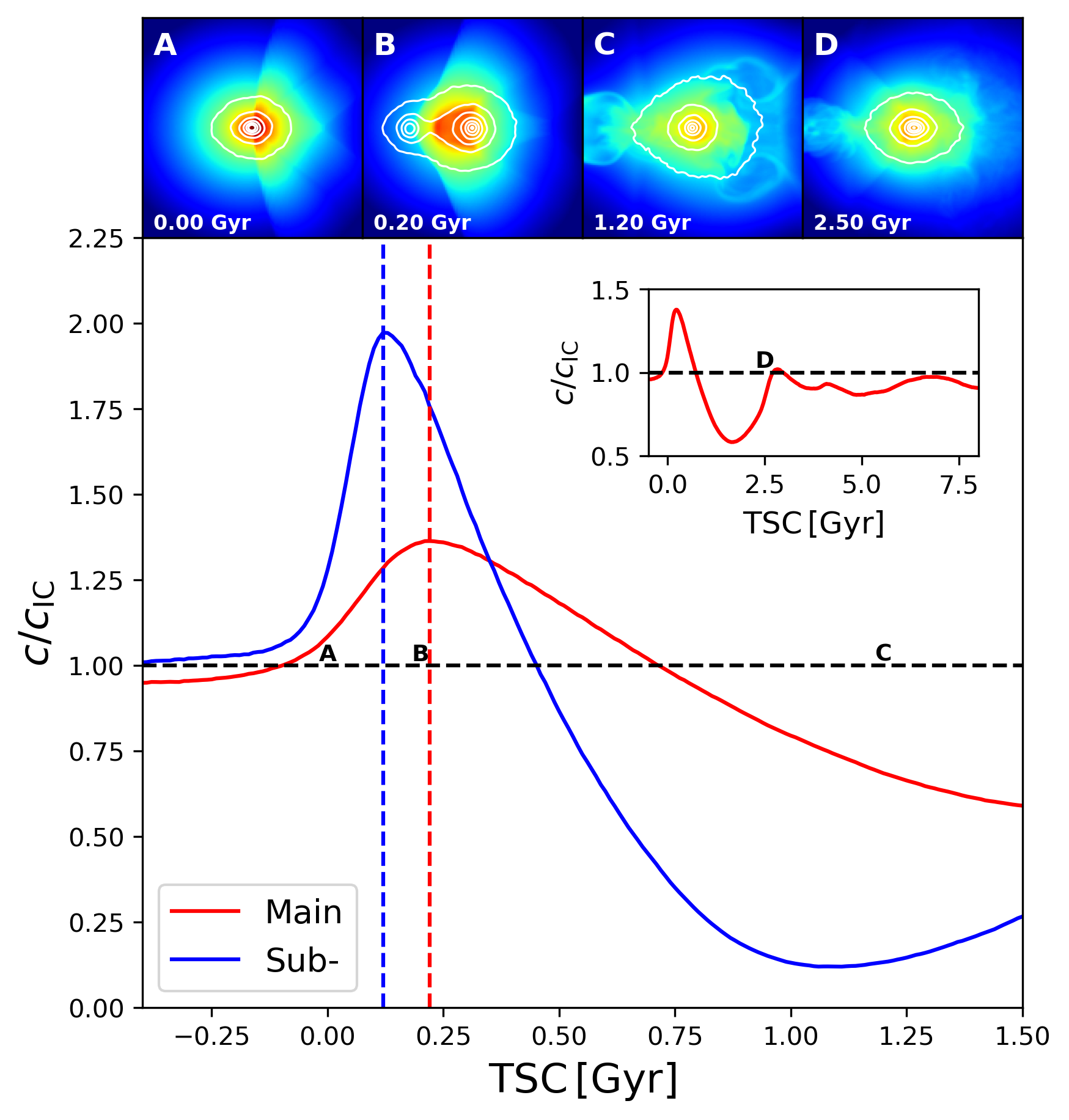}
    \caption{ 
    Time evolution of the concentration. We scale the concentration parameters $c(t)$ of the main (red) and the sub-clusters (blue) with their initial values $c_{\rm IC}$.  
    Dashed vertical lines depict the epochs of the maximum values. The main (sub-) cluster's concentration
    reaches its maximum at TSC$\sim0.23\rm~Gyrs$ ($\mytilde0.13\rm~Gyrs$), which is greater than its initial value by $\mytilde50\%$ ($\mytilde100\%$).
    The four top panels show the projected X-ray map at the first closest passage ($0$~Gyrs, A), 
    the main cluster's maximum concentration
    ($\mytilde0.2\rm~Gyrs$, B), the first apocenter ($\mytilde1.2\rm~Gyrs$, C), and the second closest passage ($\mytilde2.5\rm~Gyrs$, D). White contours represent the projected mass densities.
    The inset plot displays the time evolution of the main cluster's concentration 
    in the large (TSC$<8\rm~Gyrs$) time interval, which shows that
    the main cluster recovers its initial concentration at TSC$\sim2.5\rm~Gyrs$ near the second passage (D).}
    \label{fig:Concentration_evolution}
\end{figure}

\section{Evolution of Concentration} \label{sec:res}

Figure \ref{fig:Concentration_evolution} presents the time evolution of the concentration of the main cluster (red) and the sub-cluster (blue) in the reference run. 
As with the $M_{200}$ investigation, we normalize the concentration with its initial (pre-merger) value. 
The halo concentration significantly evolves with time since the collision. 
Before the first encounter, similar to the $M_{200}$ evolution, the concentration roughly stays consistent with its initial value. 
Shortly after the collision, we observe a rapid increase in concentration for both clusters,
which is followed by a gradual decrease.
This behavior is consistent with the findings of \citet{2022MNRAS.509.1201C}.
At the maximum, the concentration values of the main cluster and the sub-cluster are a factor of $\mytilde1.5$ and $\mytilde2.0$ greater than their initial values, respectively. 
The concentration changes faster in the sub-cluster; for the main (sub) cluster, it takes $\mytilde0.23\rm~Gyrs$ ($\mytilde0.13\rm~Gyrs$) since the first encounter to reach the maximum. 
Hereafter, we refer to the TSC for the concentration to reach its maximum as $t_{\rm peak}$.  

In the later phase of the merger, the concentration decreases below its initial value, which indicates that the halo has become more diffuse than in the pre-merger state.
The concentration of the main cluster reaches its minimum value ($\rm TSC\sim1.7\rm~Gyrs$, $\mytilde60\%$ of its initial value) during the second infall (see the subplot of Figure \ref{fig:Concentration_evolution}).
The main cluster concentration recovers its initial value at around the second closest passage ($\mytilde2.5\rm~Gyrs$), which is followed by small oscillations due to the repetitive passages of the sub-cluster.

This evolutionary trend of the concentration can be explained by a halo contraction.
As mentioned above, the halo contracts because of the increased gravitational field during the first core passage\citep[e.g.,][]{2012MNRAS.419.1338R,2022MNRAS.509.1201C}, which increases the concentration.
Physically, a larger concentration change is expected in the sub-cluster, as it experiences a more significant relative increase in gravity.

The halo contraction can also explain the difference in $t_{\rm peak}$.
When we regard the halo contraction as the collective behavior of the modified orbits of dark matter particles, the concentration should peak at half the dynamical time $t_{\rm dyn}$, which dark matter particles need to reach the center.
We can estimate $t_{\rm dyn}$ using the mass within the scale radius at the closest passage.
Then, $t_{\rm dyn}$ becomes $\mytilde0.6\,(0.3)\rm~Gyrs$ for the main (sub-) cluster.
This gives a $t_{\rm dyn}/t_{\rm peak}$ ratio of $\mytilde0.4$ in both clusters, 
which implies that the difference in $t_{\rm peak}$ can be attributed to the different orbital time of the two halos.
The actual ratio is smaller than 0.5 because the matter beyond the scale radius contracts and contributes to the enhancement of the gravitational field in practice.
This result is consistent with the analysis of \citet{2021arXiv211204926R}, which shows that a ``universal evolution" is obtained when the time is normalized with the dynamical time\footnote{We note that \citet{2021arXiv211204926R} used halo sparsity \citep{2014MNRAS.437.2328B} to quantify the compactness of halos.}.

\begin{figure*}
    \centering
	\includegraphics[width=2\columnwidth]{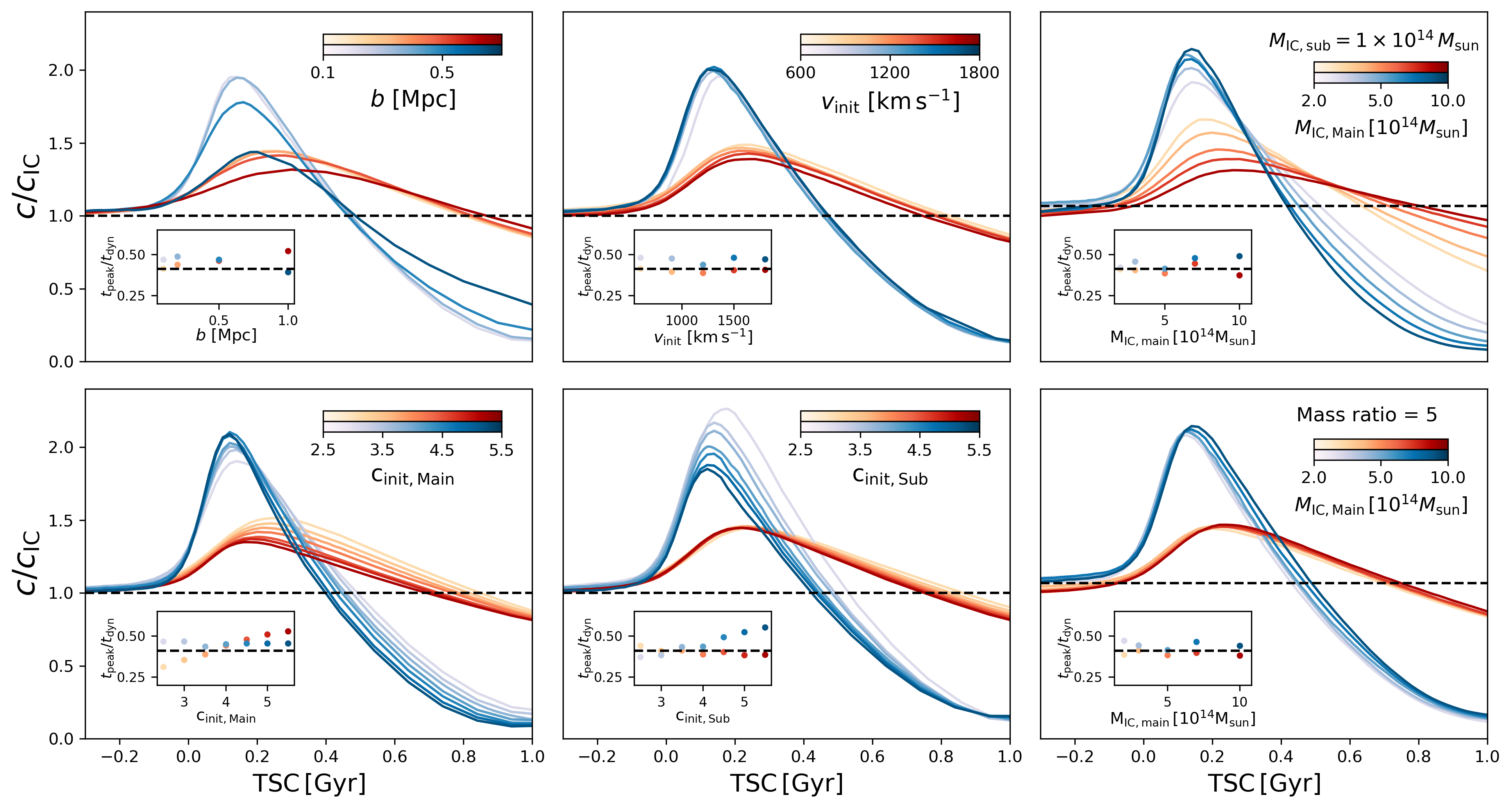}
    \caption{Time evolution of concentration during cluster mergers 
    for different collision parameters. 
    We examine the dependence of the impact parameter ($b$, top left), initial infall velocity ($v_{\rm ini}$, top middle),
    initial concentration of the main cluster ($c_{\rm init,\,Main}$, bottom left), initial concentration of the sub-cluster ($c_{\rm init,\,Sub}$, bottom middle), and main cluster mass (right column).
    The main cluster mass varies from $M_{\rm IC, Main}=2\times10^{14}M_{\odot}$ to $10^{15}M_{\odot}$ while fixing the sub-cluster mass to $M_{\rm IC, Sub}=10^{14}M_{\odot}$ (top right) or fixing the mass ratio to $5:1$ (bottom right).
    The darker color represents a larger collision parameter explored in each panel.
    The inset plot shows the $t_{\rm peak}/t_{\rm dyn}$ ratio, and the dashed horizontal line marks the ratio 0.4 from the reference run.
    The variation of the concentration evolution pattern
    can be explained by the difference in details of the halo contraction.
    The time evolution of concentration is independent of the halo mass if the mass ratio is fixed.}
    \label{fig:collision_paramter}
\end{figure*}

\subsection{Dependence on Collision Parameters} \label{sec:coll}
We show the dependence of the concentration variation on collision parameters in Figure \ref{fig:collision_paramter}.
In both clusters, a larger impact parameter results in a smaller peak value, which happens at a later epoch (top left).
The dependence on the initial velocity is insignificant (top middle) within the range explored here. 
A cluster with a higher pre-merger concentration leads to faster evolution, a lower amplitude of the concentration variation, and higher amplitude of the concentration variation in the companion cluster (bottom left and bottom right).
%When a cluster has a higher pre-merger concentration, it leads to faster evolution for both clusters, lower amplitude of the  concentration variation in the cluster, and higher amplitude of the concentration variation in the companion cluster (bottom left and bottom right).

The dependence on the collision parameters can be explained with the same halo-contraction-based reasoning mentioned in the discussion of the reference run.
Faster evolution with a larger concentration change is expected whenever the revised collision parameter provides a condition for an enhanced gravitational interaction at the closest passage. 
For example, a larger impact parameter lessens the interaction and thus leads to a smaller and slower concentration change.
A higher initial concentration, which gives a larger mass within the fixed radius, imparts a stronger gravitational force to the companion cluster.

As mentioned earlier, variation in the initial velocity shows a negligible impact on the time evolution of concentration. 
This is because the collision velocities are insensitive ($\lesssim10\%$) to the initial velocity change explored in the current study ($600-1800$~km~s$^{-1}$); even if we assume a free fall ($v_{\rm init}=0$), the resulting collision velocity is close to that in the $v_{\rm init}=600\rm~km\,s^{-1}$ case.
In summary, at the given mass, the initial velocity has an insignificant impact on the collision velocity and the time evolution of concentration.

The right panels of Figure \ref{fig:collision_paramter} shows the concentration evolution for different masses and mass ratios.
When the sub-cluster mass is fixed (top right panel), the amplitude of the main (sub-) cluster concentration variation decreases (increases) for increasing the main cluster mass. 
Again, this trend can be explained by the halo contraction argument because
the gravitational impact on the sub-cluster increases with the main cluster mass while the impact on the main cluster itself decreases.
As a result, a larger main cluster mass increases (decreases) the amplitude of the sub- (main) cluster concentration variation. 

The bottom right panel of Figure \ref{fig:collision_paramter} shows that the concentration evolution is independent of the cluster masses as long as the mass ratio is fixed. This is not surprising because the relative gravitational impacts also remain unchanged in this case; we verified this argument by repeating the experiment with a 2:1 mass ratio.

\begin{figure*}
    \centering
	\includegraphics[width=2\columnwidth]{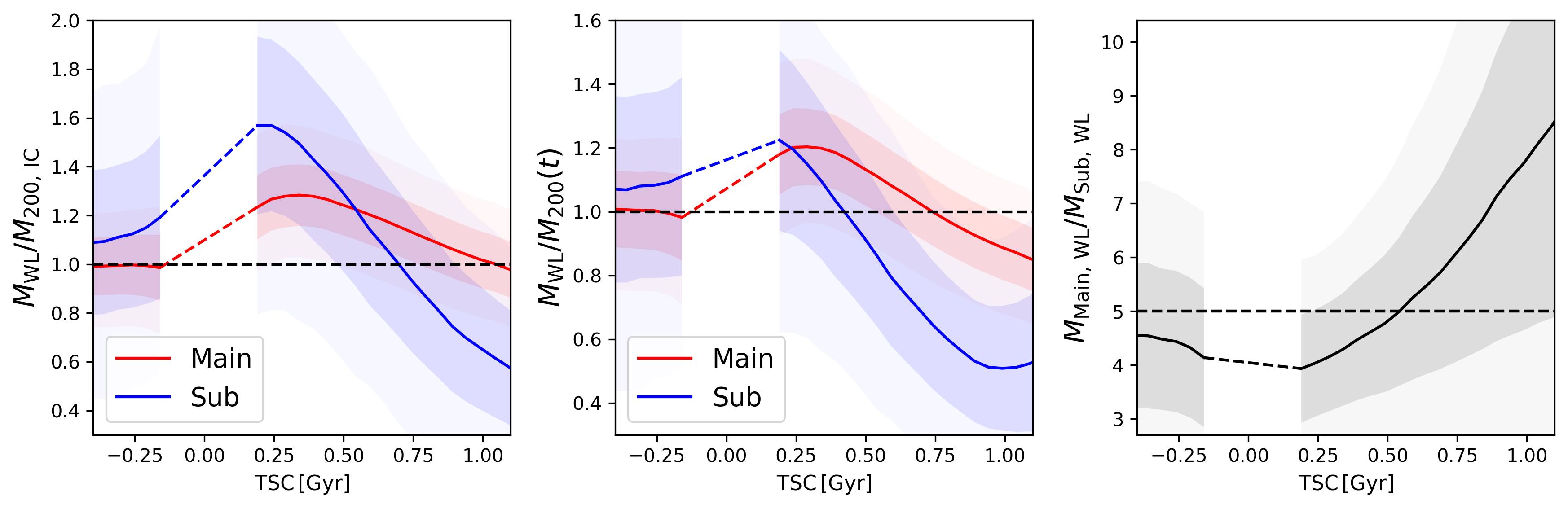}
    \caption{Time evolution of WL mass estimation in the main (red) and sub-cluster (blue) reference run. The mass of the clusters is scaled with their initial mass (left) and with $M_{200}$ measured at the current time (middle), depicting the discrepancy between WL mass and initial mass and the bias of WL mass estimation, respectively.
    The mass ratio between the estimated WL mass is presented in the right panel.
    The dark (light) shaded region depicts the WL mass uncertainty estimated with the source galaxy properties of the {\emph HST} (Subaru) observation.
    The dashed color lines indicate the flagged time zone when the cluster separation is less than $0.5\rm~Mpc$. In this range, the two-halo fitting becomes unstable.
    The evolutionary trend of the WL mass bias is similar to that of the halo concentration. 
    }
    \label{fig:WLanal}
\end{figure*}

\section{Impact on Weak-lensing Mass}
\label{sec:WLmass}

\subsection{Bias of WL Mass Estimation}\label{sec:WL_bias}

We present WL mass estimates of simulated cluster mergers at different phases in Figure \ref{fig:WLanal}. We discuss two types of mass biases. The left panel presents the bias with respect to the initial (pre-merger) mass.\footnote{Strictly speaking, this is not a bias in the usual statistical sense, as the WL analysis is not supposed to recover the pre-merger mass directly. Nevertheless, here we refer to the discrepancy as bias since typical numerical studies of merging clusters base their simulation setups on WL masses, which produce biased results.}
The middle panel presents the bias with respect to the mass at the current (observed) epoch. Since the intrinsic mass defined by the density contrast, such as $M_{200}$, indeed changes during the merger, this bias is critical when we are mainly interested in the cluster properties at the current phase.
Hereafter, we compare these mass biases with the statistical uncertainty derived assuming that the source galaxy properties follow the {\emph HST} observations. Assuming Subaru observations instead gives a consistent mass estimate, albeit with a factor of $\mytilde2$ greater mass uncertainty due to its lower source density (see dark and light regions in Figure \ref{fig:WLanal}).

The evolutionary trend of the WL mass estimate scaled by the initial mass is similar to that of concentration.
Before the collision, the WL mass is a fair representation of the initial mass for both clusters.
Then, the WL mass rapidly increases after the first passage, reaches the maximum, and decreases.
At its maximum, the WL mass is $\mytilde30\%$ ($\mytilde60\%$) higher than the initial mass of the main (sub-) cluster.
Similar to the concentration result, the sub-cluster WL mass peaks earlier ($\mytilde0.23\rm~Gyrs$) than the main cluster ($\mytilde0.34\rm~Gyrs$).
This large difference between the initial and WL mass at the observed epoch is consistent with the findings of \citet{2022MNRAS.509.1201C}. 
Interestingly, this positive correlation in the evolutionary pattern between the WL mass and halo concentration contradicts the negative correlation implied by the $M$-$c$ relation.

The middle panel of Figure \ref{fig:WLanal} shows that even when compared with the true mass at the same epoch ($M_{200}(t)$), the WL mass is still biased, and its overall variation resembles the pattern in the left panel in the sense that the bias quickly rises after the first passage, reaches the maximum, and decreases.
The largest $M_{WL}/M_{200}(t)$ ratio for the main cluster mass is $\mytilde20\%$, similar to the $M_{WL}/M_{200,IC}$ ratio. We cannot reliably estimate the equivalent ratio for the sub-cluster because the maximum bias seems to happen when the two halos are too close, and thus the simultaneous two-halo fitting becomes unstable.
In late phases, the WL mass bias shifts from overestimation to underestimation. 
The underestimation appears earlier with a more significant drop for the sub-cluster, reaching $\lesssim60$\% at TSC$\sim1.0\rm~Gyr$.

Interestingly, although the WL mass of each cluster is significantly biased during the merger, the mass ratio between the two clusters based on the WL results is relatively stable at the early (TSC$\lesssim0.6$ Gyrs) phases (right panel).
The mass ratio derived from WL increases at late (TSC$\gtrsim0.6$ Gyrs) phases, as the WL mass estimate of the sub-cluster is significantly underestimated.

\begin{figure*}
    \centering
	\includegraphics[width=2\columnwidth]{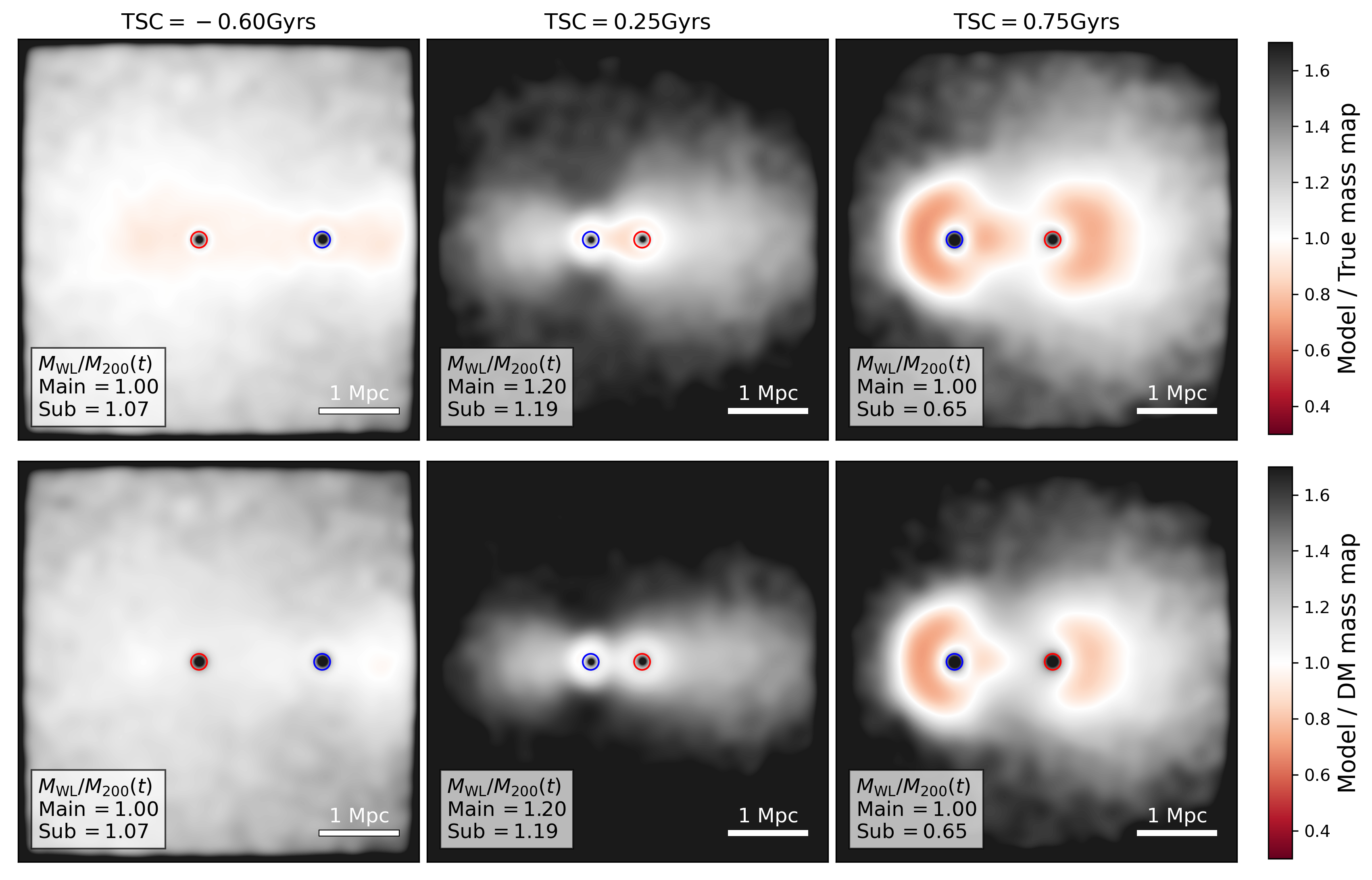}
    \caption{
    The ratio between the best-fit WL model and the simulation truth mass distribution in the reference run. 
    The top (bottom) panel compares the model mass with the total (dark matter only) mass distribution. 
    Each column displays a different merger phase from pre-merger (left), recent merger (middle), and the first apocenter passage (right).
    The $M_{\rm WL}/M_{200}(t)$ ratio at the given TSC is annotated. 
    Colored circles depict the central $100\rm~kpc$ regions we masked in the model fitting.
    WL mass over-(under)estimates the mass when the halo concentration increases (decreases), which suggests that the WL mass bias is due to the departure of the halo concentration from the assumed $M-c$ relation during the cluster merger.}
    \label{fig:Frac_dif}
\end{figure*}

To trace the source of the WL mass bias, we compare the projected mass distributions from the best-fit WL model and the values taken directly from the simulation.
The top panel of Figure \ref{fig:Frac_dif} shows the result for the reference run.
The best-fit WL model provides a good description of the true mass distribution during the pre-merger phase ($\rm TSC=-0.6~Gyrs$).
When the TSC is 0.25~Gyrs, the WL result nicely recovers the central density of both halos, whereas it overestimates the density at large radii. As mentioned earlier, 
the $M_{WL}/M_{200}(t)$ ratio at this epoch is $\mytilde1.2$ ($\mytilde1.1$) for the main (sub-) cluster.
This mass overestimation is attributed to the departure from the $M$-$c$ relation due to the merger-induced halo contraction; the main (sub-) cluster concentration increases by $\mytilde40$\% ($\mytilde70$\%) as shown in Figure~\ref{fig:Concentration_evolution}.
The increased concentration enhances the WL signal near the cluster center, which dominates the overall WL S/N.
Therefore, when we fit an NFW profile to the signal with an $M$-$c$ relation, a combination of low-concentration and high-mass results.

The mass bias at $\rm TSC=0.75~Gyrs$ can be explained by a similar logic.
In Figure \ref{fig:WLanal}, we saw that at this epoch, the WL analysis recovers the main cluster mass within 10\%, but significantly ($\lesssim60$\%) underestimates the sub-cluster mass. 
The WL mass estimate of the main cluster is close to the truth because the concentration at $\rm TSC=0.75~Gyrs$ is also close to the initial value (Figure~\ref{fig:Concentration_evolution}). On the other hand, the sub-cluster concentration drops by $\mytilde70\%$, which is a significant departure from the $M$-$c$ relation.
This suggests that the halo concentration is a critical parameter for accurate mass estimation during the post-merger.
One may argue that an NFW model might not adequately describe the mass distribution at a late merger phase because of the asymmetric mass distribution.
Nevertheless, the main cluster's $M_{200}$ is reproduced from the mock WL analysis when the halo concentration returns to the initial value (i.e., when the $M$-$c$ relation holds).

The model-to-true mass ratio in Figure \ref{fig:Frac_dif} is large near the edges. This is because the mass distribution has contracted after the collision, and the density profile becomes steeper than the NFW model prediction ($\propto r^{-3}$). Although these regions present a large ratio, they do not contribute significantly to the total mass. The WL signal from this region is small.

The bottom panel of Figure \ref{fig:Frac_dif} displays the ratio of the WL model mass to the DM-only mass. This ratio of the WL model mass to the DM-only mass helps us to identify the bias due to the (gas) model mismatch as our WL analysis fits an NFW model to the lensing signal contributed by the total (DM+ICM) mass profile, while the ICM profile does not follow the NFW profile. Comparison between the top and bottom panels illustrates that the difference due to the model mismatch at the cluster core is substantially reduced when the WL mass is compared with the DM-only mass. 

\begin{figure*}
    \centering
	\includegraphics[width=2\columnwidth]{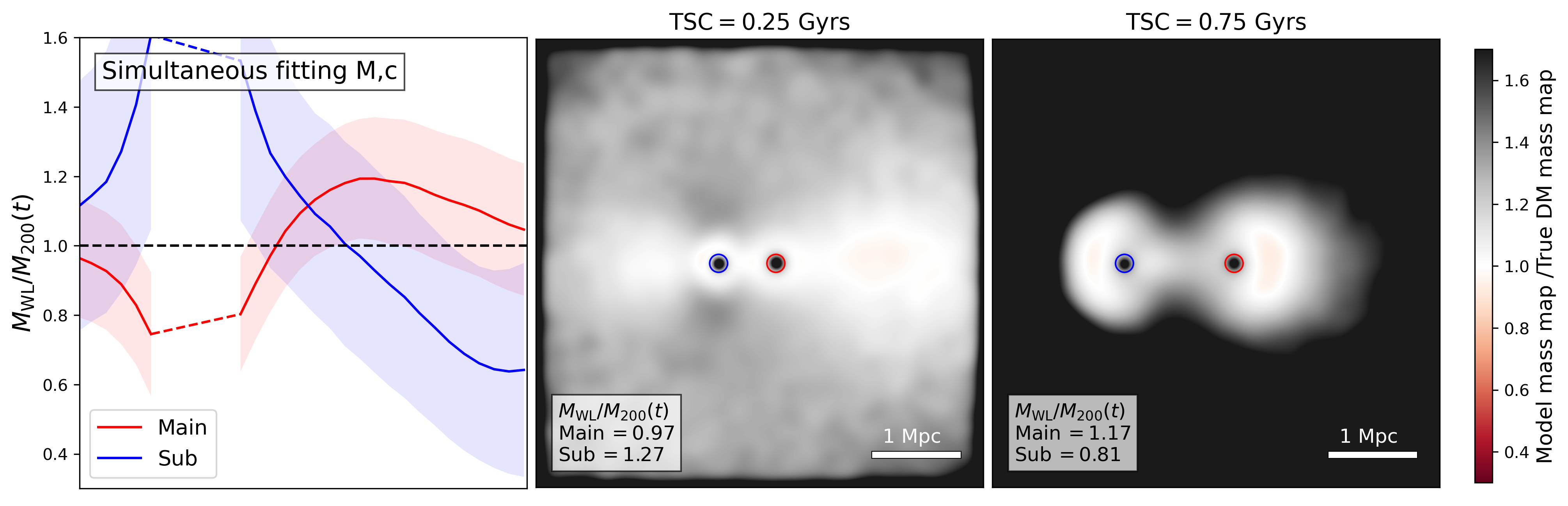}
	\includegraphics[width=2\columnwidth]{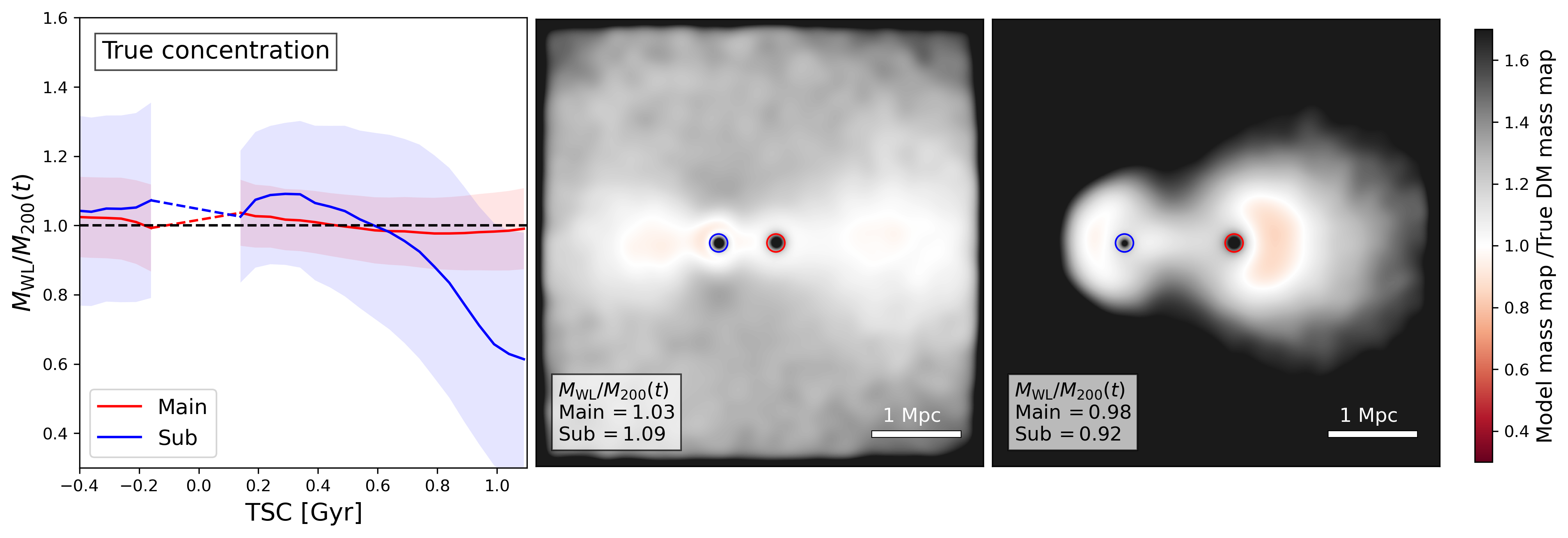}
    \caption{
    WL mass bias from simultaneously fitting mass and concentration (top) and bias from fitting mass with true concentration (bottom row). 
    The true concentration is derived with the density profile fitting. 
    The left column depicts the time evolution of WL mass bias as in Figure \ref{fig:WLanal}.
    The middle and right columns show the ratio between the WL model and the dark matter-only mass distribution as in the bottom row of Figure \ref{fig:Frac_dif}. 
    %\textcolor{red}{The WL mass estimate in the simultaneous fitting as the WL mass estimation becomes stable when the cluster separation is larger than $0.75\rm~Mpc$.}
    The mass fitting with the true concentration presents a negligible bias at TSC$\lesssim0.8\rm~Gyrs$.}
    \label{fig:wo_mcrelation}
\end{figure*}

\subsection{WL Mass Bias Reduction}

\label{sec:WLmass_and_mcrelation}
In \textsection\ref{sec:WL_bias}, we find that the WL mass bias is reduced at the epoch when the halo concentration is consistent with the $M$-$c$ prediction. We can, therefore, mitigate the WL mass bias by either not relying on the $M-c$ relation \citep[e.g.,][]{2012ApJ...758...68H,2021ApJ...918...72F, 2021ApJ...923..101K} or by obtaining a good prior knowledge of the true concentration from an independent observation \citep[e.g.,][]{2017A&A...607A..81B}.

Figure \ref{fig:wo_mcrelation} presents the WL results from our two-parameter (mass and concentration) fitting (top) and one-parameter (mass) fitting with the concentration fixed at the truth (bottom);
the true concentration is measured from the simulation data. 
When we simultaneously fit concentration and mass, the overestimation of the main cluster WL mass at $\rm TSC=0.25~Gyrs$\footnote{This is the epoch when the maximum overestimation happened in the $M-c$ relation case.} is reduced from $\mytilde 20\%$ to $\mytilde 4\%$ (top left).
The mass ratio map also shows that the simultaneous fitting provides a better description of the true mass distribution (compare the top middle panel in Figure \ref{fig:wo_mcrelation} with the top middle panel in Figure~\ref{fig:Frac_dif}).
However, this simultaneous fitting does not always improve the accuracy of the mass estimation. 
At a later phase of the merger ($\rm TSC=0.75~Gyrs$), the WL mass bias of the main cluster is greater than the result from the previous $M$-$c$ relation case. Since the level of the maximum overestimation is similar ($\mytilde20$\%), we conclude that this simultaneous fitting is not a viable solution to resolve the WL mass bias in merging clusters.

On the other hand, we find that using the accurate concentration can mitigate the WL mass bias. As shown in the bottom panel of Figure \ref{fig:wo_mcrelation}, the WL mass estimate of the main cluster is accurate within $\mytilde3$\% throughout the merger. For the sub-cluster, the accuracy is $\lesssim10$\% at $\rm TSC\lesssim0.8~Gyrs$.

This result suggests that WL can be an unbiased mass estimator in merging clusters with the aid of the true concentration. 
The halo concentration value can independently be determined with different observational tracers, such as the distribution of cluster galaxies \citep[e.g.,][]{2005ApJ...618..557N,2017A&A...607A..81B} or strong lensing \citep[e.g.,][]{2012MNRAS.420.3213O,2014MNRAS.440.1899G}.
In particular, we expect strong lensing to be a suitable solution in recent massive mergers as the merger-boosted concentration will enhance the strong lensing signals \citep[][]{2008A&A...486...35F}.

\begin{figure*}
    \centering
	\includegraphics[width=1.7\columnwidth]{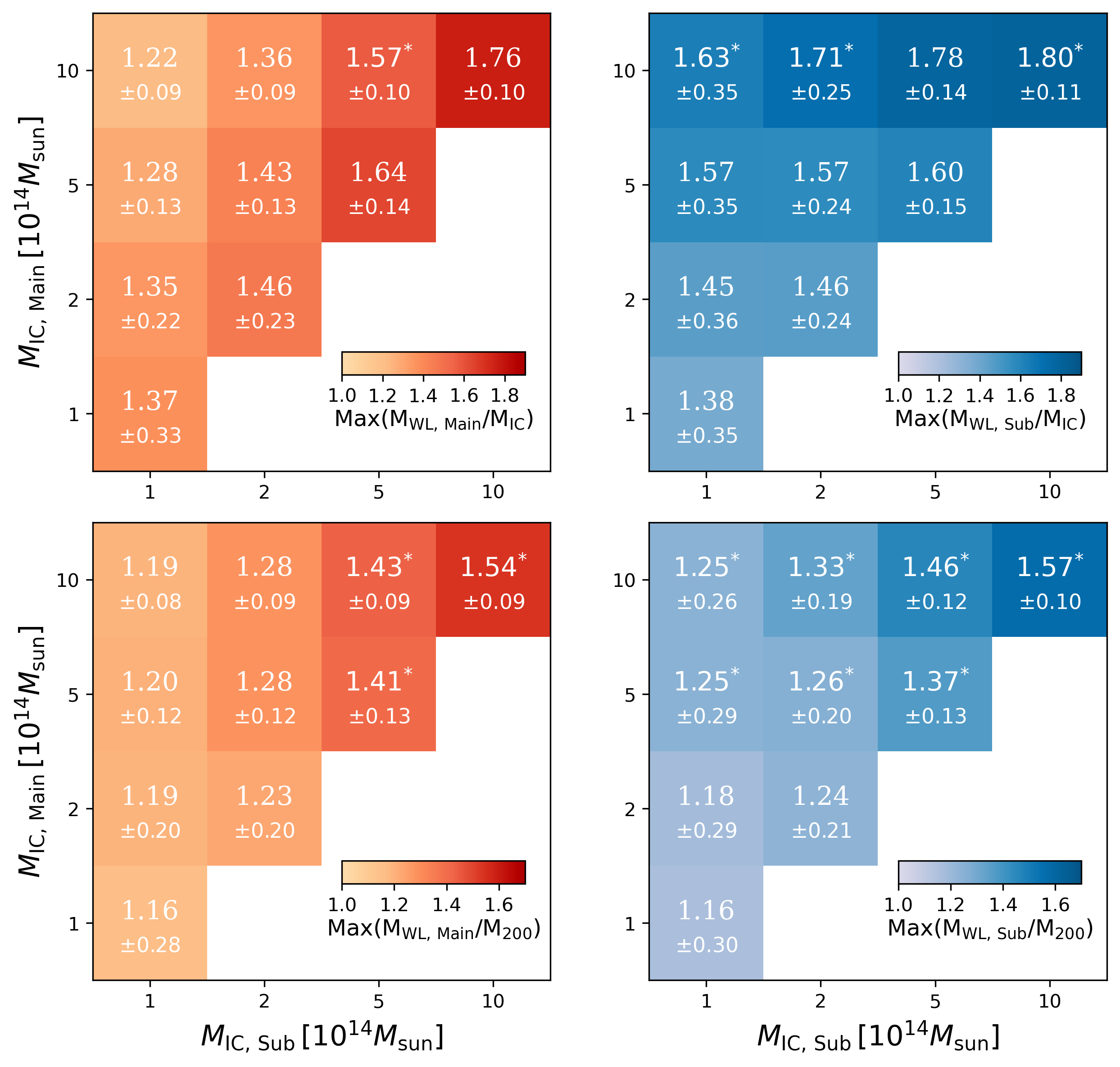}
    \caption{
    Maximum $M_{\rm WL}/M_{\rm IC}$ (top) and $M_{\rm WL}/M_{200}(t)$ ratio (bottom) of the cluster mergers with different initial masses.
    The initial mass of the main and the sub-clusters are annotated in the axis. 
    Asterisks mark the cases whose fitting is unstable because of the insufficient halo separation.
    The WL mass bias increases in both clusters with the increasing sub-cluster mass.
    }
    \label{fig:WLbias_difmass}
\end{figure*}

\subsection{Mass Dependence of WL Mass Bias}\label{sec:WL_in_different_mergers}

Figure \ref{fig:WLbias_difmass} shows how the WL mass bias changes as we vary the initial masses. Since the true and WL masses change as the merger progresses, we use the maximum ratio with respect to the initial-true (top) and current-true (bottom) masses to quantify the dependence. Since our snapshot time resolution is finite, the maximum value chosen from the snapshot data is always underestimated.
Thus, we refine the maximum ratio by fitting a second-order polynomial to the three values within $20\rm~Myrs$ from the tentative maximum. Note that sometimes, the epoch where the maximum occurs likely resides within the time interval where our simultaneous fitting is unstable (i.e., the halo separation is too small). For these cases, we mark the value with an asterisk.

With respect to the initial-true mass, the WL mass bias of both clusters increases with the companion cluster mass. For the main cluster, the mass bias decreases with its mass, whereas this self-dependence is insignificant for the sub-cluster. It is remarkable that for the 1:1 merger of two $10^{15}~M_{\sun}$ clusters, the maximum bias is nearly $\mytilde80$\%.

When the bias is measured with respect to the current-true mass, the dependence on the main cluster mass is insignificant in both clusters. 
For instance, with a $10^{14}~M_{\sun}$ sub-cluster, the $10^{14}-10^{15}~M_{\sun}$ variation in the main cluster mass affects the bias of the main cluster at the $\mytilde4\%$ level, which is $\mytilde10\%$ of the uncertainty. This insensitivity can be understood from the evolution of the current-true mass $M_{200}(t)$.
The main (sub-) cluster shows an increased time variation of $M_{200}(t)$ for decreasing (increasing) main cluster mass and increasing (decreasing) sub-cluster mass, which is similar to the mass dependence of $M_{WL}$ that was discussed above.
Therefore, the dependence on the main cluster mass is cancelled in the evaluation of  $M_{WL}/M_{200}(t)$.
We note that a similar mass dependence is observed when we simultaneously fit the mass and concentration (as in \textsection\ref{sec:WLmass_and_mcrelation}).

Figure \ref{fig:WLbias_difmass} presents significant implications for interpreting various merger observations. For instance, WL mass bias will be small or negligible between low-mass ($\mytilde10^{14}~M_{\odot}$) systems or in a massive cluster with a group-size collider. %a minor merger, which involves a group-size collider. 
Even at its maximum, the level of bias is insignificant, compared to the statistical uncertainty of the WL mass \citep[\(\mytilde50\%\),][]{2021ApJ...918...72F}.
In contrast, the bias is more significant in massive mergers.
The difference between the WL mass estimate and the current-true mass can reach up to $\mytilde40\%$ in the collision of two moderately massive ($5\times10^{14}~M_{\odot}$) clusters.
As this bias is significant relative to the statistical uncertainty \citep[\(\lesssim30\%\),][]{2021ApJ...923..101K}, the bias correction is required.

Another important takeaway from this investigation is that the maximum bias occurs in an early phase of the merger ($0.2-0.4$~Gyrs after the first encounter), which includes several merger cases reported in the literature.
Thus, some existing WL studies may have overestimated the masses, mainly when the cluster selection is based on their prominent merger features.

\section{Application to Observations}
\label{sec:obs}
Our mock WL analysis shows that a WL mass can be significantly overestimated in a massive recent merger. 
The large discrepancy between the WL estimate and the true pre-merger mass significantly impacts our interpretation of the merger, including the collision scenario reconstruction.
In this section, we investigate the impact for
the three observational cases: A2034, MACS1752, and ZwCl1856, which
have been recently identified as exemplary binary mergers with a near head-on collision.

\subsection{Review of Observations and Simulation Setup}

A2034 is a dissociative merger where its merger shock is detected with the X-ray feature discontinuity \citep{2014ApJ...780..163O} whereas
MACS1752 and ZWCL1856 are the mergers with double radio relics \citep{2012MNRAS.426...40B,2014MNRAS.444.3130D}. A radio relic is a Mpc-scale extended diffuse radio emission feature, which often traces the merger shock front as seen in X-rays \citep[e.g.,][]{2012A&ARv..20...54F}.
Radio relics come in various shapes due to the diversity in merger configurations and in cluster environments.
The double radio relic systems in MACS1752 and ZWCL1856
are remarkably symmetric and highly consistent with the hypothesized collision axis and the locations of the merger shocks, as predicted by numerical simulations \citep[e.g.,][]{Ha2018}. 

The A2034 WL analysis was performed by \citet{Kyle_Thesis}. 
\citet{2021ApJ...918...72F} presented WL studies of MACS1752 and ZWCL1856.
Based on the source density of $\gtrsim20$ galaxies $\rm arcmin^{-2}$, these WL studies identified two halos in each system, and their masses were derived by NFW profile fitting with the $M-c$ relation, as is done in our mock WL analysis.
\cite{Kyle_Thesis}
shows that A2034 is a $\mytilde4:1$ merger between $3.6\pm0.7$ and $0.9\pm0.5\times10^{14}\,M_{\odot}$ halos.
According to \citet{2021ApJ...918...72F}, MACS1752 (ZWCL1856) is an equal-mass merger with halos of mass $4.7\pm0.8$ and $3.6\pm0.7\times10^{14}\,M_{\odot}$ ($1.2\pm0.6$ and $1.0\pm0.6\times10^{14}\,M_{\odot}$).
Readers are referred to \citet{Kyle_Thesis} and \citet{2021ApJ...918...72F} for the analysis details.

For simplicity, we describe the three systems with the combinations of $5\times10^{14}~M_{\odot}$ and $1\times10^{14}~M_{\odot}$ halos. Specifically, we represent A2034 with the head-on collision  between $5\times10^{14}~M_{\odot}$ and $1\times10^{14}~M_{\odot}$ halos (i.e., reference run) whereas the MACS1752 (ZwCl1856) analog is
a merger between two $5(1)\times10^{14}~M_{\odot}$ halos.
X-ray observations clearly show that both MACS1752 and ZwCl1856 maintain their gas cores.
Thus, we set a non-zero impact parameter of $0.6\,(0.3)\rm~Mpc$ for MACS1752 (ZwCl1856) so that the gas core survives after the first passage.
One can fine-tune the simulation setup so that our mock WL mass closely matches the reported values in \cite{2021ApJ...918...72F} and \citet{Kyle_Thesis}. However, we do not attempt to do it here since our primary goal is to estimate the level of bias to the first order.

\subsubsection{Shock-based Time Estimation}
\label{sec:shock_based_time}

Before we discuss the time evolution of the WL mass bias, we introduce the shock-based time $\tau_{\rm sh}$:
\begin{equation}
    \tau_{\rm sh}=\frac{d_{\rm sh}}{V_{200}},
\end{equation}
\noindent
where $d_{\rm sh}$ and $V_{200}$ are the shock-to-shock distance and the virial velocity based on the summation of the two $M_{200}$ values.
We assume that the merger is happening on the plane of the sky. Thus, $d_{\rm sh}$ and $\tau_{\rm sh}$ are the lower limits.
Since we can identify the shock locations in both observation and simulation, $\tau_{\rm sh}$ is a convenient measure of the TSC applicable to both observation and simulation.

For the observations of MACS1752 and ZwCl1856, we adopt the distance between the two radio relics as $d_{\rm sh}$. Since the A2034 shock was detected only on one side so far, we use the distance between the shock and the main halo as the time proxy.

For the simulation analogs, we create a gas velocity divergence map and find the locations of the minimum \citep[maximum convergence, ][]{Ryu2003,Skillman2008}. 
In the case of the MACS1752 and ZwCl1856 analogs, we adopt $\tau_{sh}$ as the distance between the two local minima.
For A2034, we use the distance between the main halo-side shock and the main halo to match the observational procedure.

\begin{figure}
    \centering
	\includegraphics[width=\columnwidth]{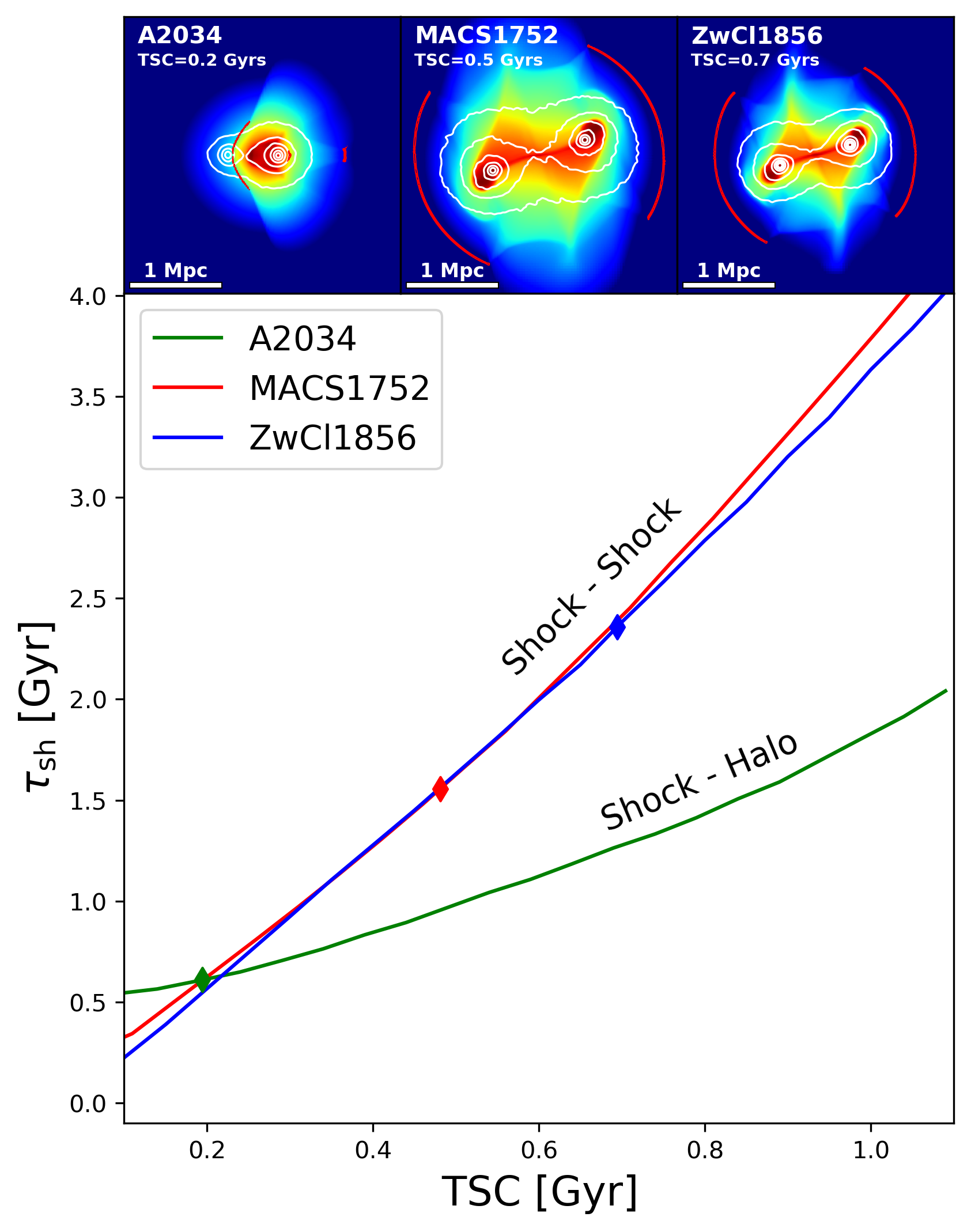}
    \caption{Relation between the TSC and the time derived with shock separation ($\tau_{\rm sh}$) in the simulation analogs. The $\tau_{\rm sh}$ is derived by dividing shock to halo (A2034) or shock to shock distance (MACS1752, ZWCL1856) with the virial velocity ($V_{200}$) that is estimated with sum of WL mass estimate ($M_{\rm WL,\,Main}+M_{\rm WL,\,Sub}$).
    Diamond markers mark the observed quantities, and we read the TSC as $\mytilde0.2$, $\mytilde0.5$, and $\mytilde0.7~\rm Gyrs$ for A2034, MACS1752, and ZWCL1856, respectively. 
    The top panels show the X-ray surface brightness map overlaid with the mass (white) and the velocity divergence contours (i.e., shock contours, red) at the estimated TSC.
    Characteristic features of observed mergers are reproduced at the estimated TSC.}
    \label{fig:MergerShock}
\end{figure}

Figure~\ref{fig:MergerShock} displays the relation between our shock-based time measure $\tau_{sh}$ and the true time (TSC) from the simulation analogs. It is important to note that the relation is quasi-linear for the time interval shown here. This is because $V_{200}$ is nearly linear with time, although the mass sum is significantly nonlinear.
The $\tau_{sh}-TSC$ linearity demonstrates that $\tau_{sh}$ can serve as a valuable measure of the TSC, as its true value is not directly accessible from observations.

We can read off the TSC values for the three clusters with Figure~\ref{fig:MergerShock}. They are estimated to be $\mytilde0.2$, $\mytilde0.5$, and $\mytilde0.7\rm~Gyrs$ for A2034, MACS1752, and ZWCL1856, respectively.
These estimates are comparable to those in the previous studies. In the case of A2034, our estimate $\mytilde0.2$~Gyrs is slightly lower (earlier merger phase) than the literature values.
% share the result of https://ui.adsabs.harvard.edu/abs/2021MNRAS.500.1858M/abstract
\citet{2014ApJ...780..163O}, \citet{2021MNRAS.500.1858M}, and \citet{2018MNRAS.481.1097M}
quoted $\mytilde0.3\rm~Gyrs$, $\mytilde0.3\rm~Gyrs$, and $\mytilde0.5\rm~Gyrs$
based on the X-ray shock velocity, the offset between X-ray and DM peaks, and the spectroscopic and weak-lensing analysis, respectively.
The difference mainly comes from the viewing angle since our plane-of-the-sky-collision assumption provides a lower limit. We will discuss the impact of the viewing angle on the WL mass bias further in \textsection \ref{sec:viewing_angle}.
Nevertheless, the key characteristics, such as the X-ray front over the sub-cluster in A2034 or the X-ray tails of MACS1752, are well-reproduced at $\tau_{\rm sh}$ (top panel of Figure \ref{fig:MergerShock}). This reassures that $\tau_{\rm sh}$ is a reasonable proxy for the TSC for the first-order interpretation of the observations.

\subsection{WL Mass Bias at Observed Epoch}

\label{sec:bias_in_obs}

Based on the WL bias vs. TSC relation from the simulation analogs, we can determine the WL bias for the three systems. With respect to the current mass,
the WL masses are likely to be overestimated by $\mytilde20$\% and $\mytilde35$\% for A2034 and MACS1752, respectively, whereas a $\mytilde10\%$ underestimation is expected for ZwCl1856.
For A2034, the bias is close to the maximum since the estimated TSC$\sim0.2$~Gyrs is near the epoch of the maximum bias (TSC$\sim0.25$~Gyrs, see the middle panel of Figure~\ref{fig:WLanal}).
In the case of MACS1752, the WL bias is estimated to be greater than that of A2034 mainly because of the higher system mass. Finally, we expect a mass underestimation for ZWCL1856 due to its late merger phase.
As mentioned earlier, the mass bias in MACS1752 is comparable to the statistical uncertainty.

\subsection{WL Mass Bias at Pre-merger Epoch and Method for Correction}
\label{sec:WLtoPre}
Regarding following up merger observations with simulations,
priority should be given to accurate mass estimation at the pre-merger epoch.
We find that the WL mass estimate at the observed epoch is larger than the initial mass at the pre-merger epoch in all three cases. We expect $\mytilde30\%$, $\mytilde70\%$, and $\mytilde20\%$ overestimations for the main cluster mass of A2034, MACS1752, and ZwCl1856, respectively.
%This implies that we should decrease the pre-merger mass to reproduce the observed WL mass estimate. 

\begin{figure}
    \centering
	\includegraphics[width=\columnwidth]{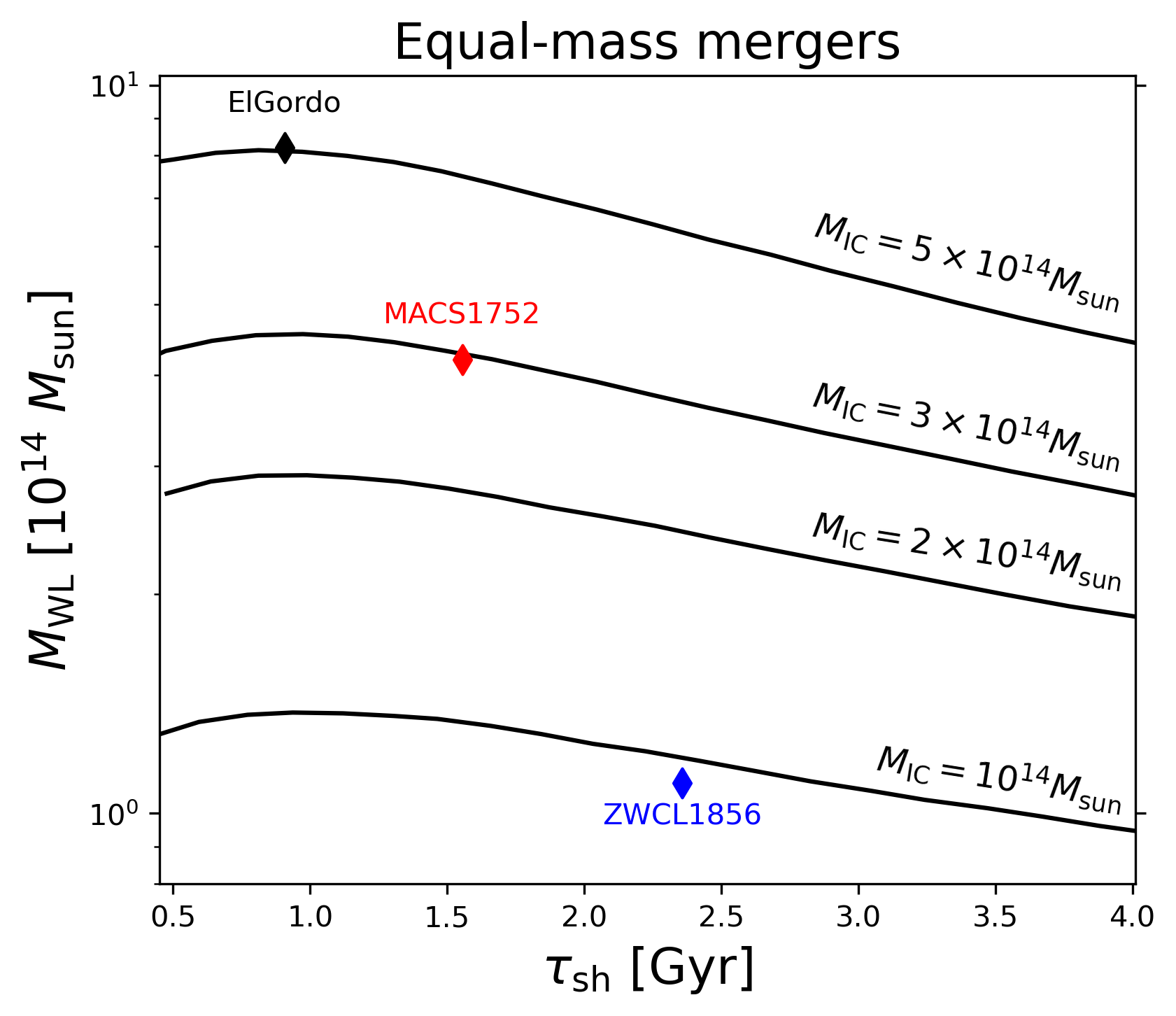}
    \caption{Time evolution of WL mass estimate for equal-mass mergers. 
    Four equal mass mergers with different initial masses are presented, and the initial mass of a single halo $M_{\rm IC}$ is annotated over the curve.
    Diamond markers mark the observed quantities of the three cluster mergers.
    The observed WL mass estimate can be explained with a collision between less-massive clusters (i.e., $M_{\rm WL}\gtrsim M_{\rm IC}$).}
    \label{fig:WLmasstoIC}
\end{figure}

Our scheme for the derivation of the bias-corrected pre-merger mass is as follows.
We utilize the notion that the WL analysis can reproduce the initial mass ratio although each mass is biased (\textsection \ref{sec:WL_bias}).
First, we generate a series of merger simulations
with the observed mass ratio while varying the total system mass. This allows us to map out the time evolution of the WL masses for different total masses. Then, we find
the simulation set whose predicted WL mass matches the observation. The initial mass of the matching simulation set can be adopted as the bias-corrected pre-merger mass.

Figure \ref{fig:WLmasstoIC} illustrates our scheme for the equal-mass merger samples: MACSJ1752, ZWCL1856, and ACT J0102-4915.
For MACS1752, we find that the observed WL mass estimate $\mytilde4.1\times10^{14}~M_{\odot}$ at $\tau_{sh}\sim1.6$~Gyrs (TSC$\sim0.5$~Gyrs)
is reproduced from the collision of two $3\times10^{14}~M_{\odot}$ halos.
Similarly, we can explain the observed WL mass of ZWCL1856 ($\mytilde1.1\times10^{14}~M_{\odot}$) with an equal mass merger between two $10^{14}~M_{\odot}$ halos.
The ACT J0102-4915 system, nicknamed ``El Gordo", is one of the most massive mergers to date \citep{2021ApJ...923..101K}. 
Figure~\ref{fig:WLmasstoIC} suggests that the current observed WL masses ($\mytilde10^{15}$ and $\mytilde6.5\times10^{14}~M_{\sun}$) can result from the collision of two $\mytilde5\times10^{14}~M_{\odot}$ halos when the merger is observed at TSC$\sim0.3$~Gyrs.
The expected pre-merger mass of El Gordo is still massive, but it is $\mytilde40\%$ less massive than the WL mass estimate \citep{2021ApJ...923..101K}.
The estimated TSC of El Gordo is a factor of $\mytilde2$ larger than the result from \citet{2015ApJ...813..129Z}, who used a set of idealized collisions between two $10^{15}M_{\odot}$ halos to represent the cluster merger\footnote{A returning phase (i.e., after the first apocenter passage) has been suggested by \citet{2015MNRAS.453.1531N} and \citet{2021ApJ...923..101K} based on their Monte-Carlo simulations. 
In our time estimation, this late phase is disfavored as the shock separates more than $3.5\rm~Mpcs$ when the halo starts its second infall.}.

%%%%%%%%%%%%%%%%%%%%
\section{Discussion}\label{sec:discussion}
\subsection{Dependence on Viewing Angle} 
\label{sec:viewing_angle}

\begin{figure}
    \centering
	\includegraphics[width=\columnwidth]{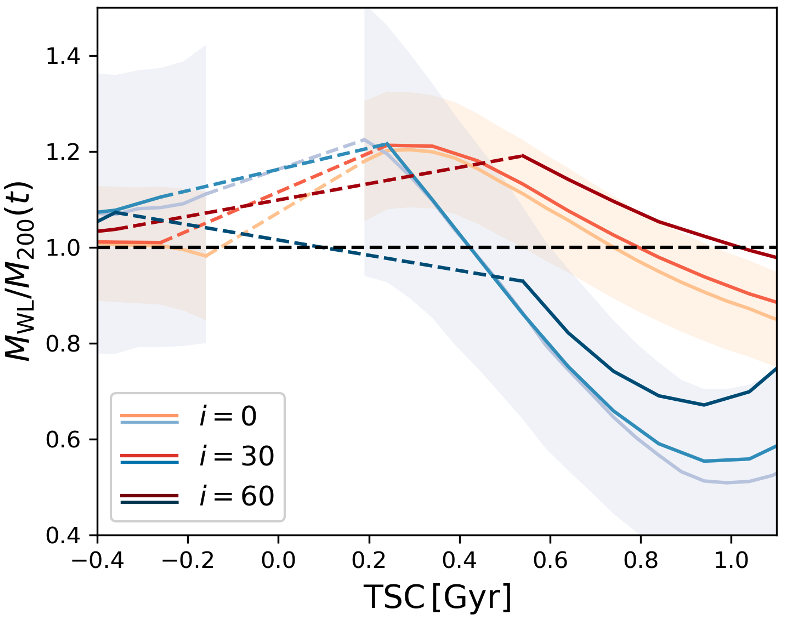}
    \caption{Time evolution of the WL mass bias with a different viewing angle ($i$). 
    Darker color represents a larger viewing angle, and a shaded region marks the uncertainty in the plane-of-the-sky collision ($i=0$). 
    The WL mass estimate marginally increases with the increasing viewing angle, but the variation is within the uncertainty.}
    \label{fig:ViewingAngle}
\end{figure}

Throughout the paper, we assumed a plane-of-the-sky merger (i.e., zero viewing angle).
This assumption might not be entirely invalid for double radio relic systems because a projection with a large viewing angle is known to hinder the identifications of radio relics \citep[see Figure 7 of][]{2013ApJ...765...21S}.
However, it is reasonable to assume that the real merger axis is somewhat tilted with respect to the plane of the sky.
This section examines the impact of non-zero viewing angles on the WL mass bias. 

Figure \ref{fig:ViewingAngle} shows that the viewing angle has a negligible impact on the time evolution of WL mass bias. 
With increasing viewing angle, the $M_{WL}/M_{200}(t)$ ratio marginally increases, yet the difference is within the uncertainty.

The WL mass bias is consistent among different viewing angles because of the model fitting.
In our mock WL analysis, the mass distribution is simultaneously fitted with the model of two projected spherical halos. 
This allows the WL analysis to measure the overlapping mass from the two halos, so the viewing angle does not significantly change the final mass estimate. 

Nevertheless, we caution that the viewing angle is still a critical parameter because the constraint on TSC using shock separation can vary with the viewing angle.
Nonetheless, we can correct the WL mass bias to the first order with the method described in \textsection \ref{sec:WLtoPre} by deprojecting the observed shock separation with the viewing angle constrained from independent observations \citep[e.g., radio polarization, ][]{1998A&A...332..395E}.
Note that our discussions on the observational examples in \textsection \ref{sec:WLtoPre} still hold because these mergers might be happening nearly in the plane of the sky \citep[][]{2018ApJ...862..160W}.

\begin{figure}
    \centering
	\includegraphics[width=\columnwidth]{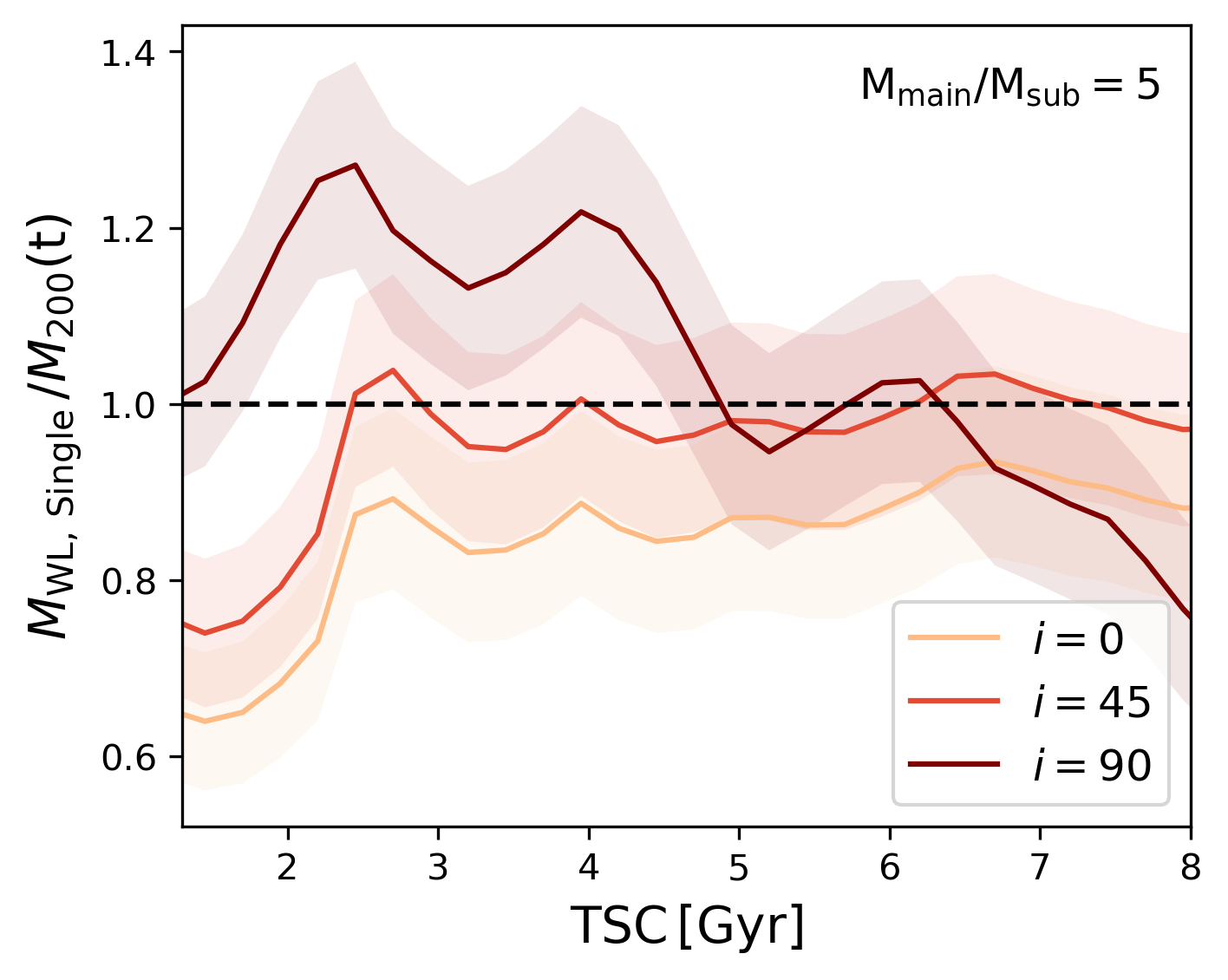}
    \caption{WL mass bias of single halo mass fitting in the reference run.
    Color represents different viewing angles. 
    The fitted mass with single halo model $M_{\rm WL,\,Single}$ is scaled with $M_{200}$, and the time axis is measured from the first closest passage. 
    The WL mass estimate increases with the viewing angle, and the mass of the line-of-sight merger ($i=90$) can be overestimated by $\mytilde20\%$ until a late merger phase (TSC$\sim4.5$~Gyrs).}
    \label{fig:Single}
\end{figure}

\subsection{WL Mass Estimate in a Late Merger Phase} \label{sec:SingleCluster}

In \textsection\ref{sec:WLmass}, we show that the WL analysis underestimates the halo mass as the clusters reach their first apocenter. 
At the same epoch, we also find that the sub-cluster quickly disrupts. 
Therefore, the observations will barely recognize the sub-cluster center as in the simulations, and so the two halo mass fitting will be unavailable in a late merger phase. 

In Figure \ref{fig:Single}, we perform the WL analysis in a late merger phase of our reference run with a single halo model. The single halo model is fitted with its center fixed on the main cluster. 
We scale the WL mass estimate, $M_{\rm WL,Single}$, with the current mass, $M_{200}(t)$, that we measured using all dark matter particles and gas mass density.

We find that the merger can have a long-lasting effect on the WL mass estimate.
In the plane-of-the-sky collision, the WL analysis moderately underestimates the current mass, yet the difference is comparable to the mass uncertainty. 
On the other hand, the WL mass estimate overestimates the mass by $\mytilde20\%$ in the line-of-sight mergers. 
This overestimation is more significant than the mass uncertainty until TSC$\sim4.5\rm~Gyrs$. 

We can explain the mass overestimation in the line-of-sight mergers with the discussions in \textsection\ref{sec:WLmass}. 
The WL analysis over(under)estimates the current mass when the halo concentration is larger (smaller) than what the $M-c$ relation predicts.
Thus, mass overestimation is expected in the line-of-sight collision as the mass distribution is compact in the center by the projection.
This result is consistent with the WL mass bias of fitting triaxial halos, where mass is overestimated when projected along the major axis \citep[e.g.,][]{2004MNRAS.350.1038C}.

An important takeaway is that an old merger event can affect the current WL mass estimate.
Clusters can experience multiple mergers in $\mytilde5$~Gyrs. 
Our experiment suggests that a collision within the past $\mytilde5$~Gyrs can bias the WL mass estimate at the current epoch. 
Moreover, the WL mass bias was comparable to the mass uncertainty in a single system.
However, the bias will become significant compared to the uncertainty of the mass averaged in a cosmological context. 
The WL mass bias throughout cosmic history is outside the scope of this work, and we plan to address it in future studies. 

\section{Conclusions} \label{sec:conclusions}
In this study, we have investigated the bias of WL mass estimation in deriving the pre-merger and the current mass of merging clusters. 
We performed a realistic mock WL analysis with observed source galaxy properties on a suite of cluster merger simulations. 
The main findings of our paper can be summarized as the followings. 

\begin{itemize}
    \item We found that the concentration of the dark halo increases after the first encounter, gradually decreases, and recovers after the second closest passage. This evolutionary trend can be explained by gravitational contraction of the halo. We found that the time evolution of concentration varies with different collision parameters and the trend was consistent with the total mass when the mass ratio is fixed. 
    
    \item In a collision between $5$ and $1\times10^{14}M_{\odot}$ clusters, the WL analysis can overestimate the current mass by up to $\mytilde20\%$ at TSC$\sim0.2\rm~Gyrs$. The current mass is underestimated in a late merger phase. Despite the bias in WL mass estimate, we found that the mass ratio between the two halos can be reproduced during the early stage of the merger (TSC$\lesssim0.6\rm~Gyrs$). 
    
    \item We found a positive correlation between the time evolution of the WL mass and that of the halo concentration. 
    Based on this correlation, we attribute the WL mass bias to the halo concentration that departs from the M-c relation during the cluster merger.
    We can control the WL mass bias at $\mytilde3\%$ level by providing the accurate concentration value using the distribution of cluster galaxies or lensing observations.
    %The WL mass bias can be reduced to $\mytilde3\%$ level when the true concentration is provided.
    %We suggest that the true concentration can be constrained by observations of galaxy distribution or strong lensing. 
    
    \item We found that the WL mass bias increases with the mass of the companion cluster.
    The difference between the WL and current mass can reach $\mytilde60\%$ in a collision between $10^{15}~M_{\odot}$ halos. 
    Based on this mass dependence, we suspect that the bias has a negligible impact on minor or low-mass cluster mergers. 
    In contrast, a proper bias correction is necessary for massive cluster mergers. 
    
    \item We estimated the WL mass bias in the observational examples of A2034, MACS1752, and ZwCl1856. 
    We demonstrated that the spatial separation of the merger shocks could provide a reliable reference for estimating TSC. 
    We found that the WL mass estimate can be over- and underestimated depending on the merger phase and the total mass. 
    
    We further derive the bias-corrected initial mass by utilizing the fact that WL analysis can reproduce the initial mass ratio.
    We found that the difference between the WL mass estimate and the pre-merger mass increases with the halo mass. 
    The difference becomes $\mytilde40\%$ in $10^{15}M_{\odot}$ mass clusters, which can modify the merger history reconstructed based on the WL mass estimate. 
    
\end{itemize}

Our results suggest that the previous WL mass estimates of massive merging clusters, using the parametric model fitting, may be biased high. 
Since the bias depends on the TSC, an accurate reconstruction of merger history is essential for alleviating the WL mass bias of merging clusters.
This study was limited to merging clusters inside an isolated box.
For a more direct comparison with the observations, future studies need to examine the WL mass bias in a cosmological volume.

\hfill \break
M. J. Jee acknowledges support for the current research from the National Research Foundation (NRF) of Korea under the program 2022R1A2C1003130. JAZ acknowledges support from the Chandra X-ray Center, operated by the Smithsonian Astrophysical Observatory for and on behalf of NASA under contract NAS8-03060.

\bibliography{reference}

\end{document}